\documentclass{aa}

\usepackage{graphicx}
\usepackage{txfonts}
\usepackage{hyperref}
\usepackage[noabbrev]{cleveref}
\hypersetup{draft}

\begin{document} 

   \title{{The} polarization-encoded self-coherent camera}
   
   \author{S.P. Bos}
   \institute{Leiden Observatory, Leiden University, P.O. Box 9513, 2300 RA Leiden, The Netherlands \\     
               \email{stevenbos@strw.leidenuniv.nl}
             }

   \date{Received October 1, 2020; accepted December 15, 2020}

 
  \abstract
   {The exploration of circumstellar environments by means of direct imaging to search for Earth-like exoplanets is one of the challenges of modern astronomy.
    One of the current limitations are evolving non-common path aberrations (NCPA) that originate from optics downstream of the main wavefront sensor. 
    Measuring these NCPA with the science camera during observations is the preferred solution for minimizing the non-common path and maximizing the science duty cycle. 
    The self-coherent camera (SCC) is an integrated coronagraph and focal-plane wavefront sensor that generates wavefront information-encoding Fizeau fringes in the focal plane by adding a reference hole (RH) in the Lyot stop.
    However, the RH is located at least 1.5 pupil diameters away from the pupil center, which requires the system to have large optic sizes and results in low photon fluxes in the RH.  
     }
   {Here, we aim to show that by featuring a polarizer in the RH and adding a polarizing beamsplitter downstream of the Lyot stop, the RH can be placed right next to the pupil.
   This greatly increases the photon flux in the RH and relaxes the requirements on the optics size due to a smaller beam footprint. 
   We refer to this new variant of the SCC as the polarization-encoded self-coherent camera (PESCC).
   }
   {We study the performance of the PESCC analytically and numerically, and compare it, where relevant, to the SCC.
   We look into the specific noise sources that are relevant for the PESCC and quantify their effect on wavefront sensing and control (WFSC). 
   }
   {
   We show analytically that the PESCC relaxes the requirements on the focal-plane sampling and spectral resolution with respect to the SCC by a factor of 2 and 3.5, respectively. 
   Furthermore, we find via our numerical simulations that the PESCC has effectively access to $\sim$16 times more photons, which improves the sensitivity of the wavefront sensing by a factor of $\sim4$. 
   We identify the need for the parameters related to the instrumental polarization and differential aberrations between the beams  to be tightly controlled -- otherwise, they limit the instrument's performance. 
   We also show that without additional measurements, the RH point-spread function (PSF) can be calibrated using PESCC images, enabling coherent differential imaging (CDI) as a contrast-enhancing post-processing technique for every observation.  
   In idealized simulations {(clear aperture, charge two vortex coronagraph, perfect DM, no noise sources other than phase and amplitude aberrations)} and in circumstances similar to those of space-based systems, we show that WFSC combined with CDI can achieve a $1\sigma$ raw contrast of $\sim3\cdot10^{-11}- 8 \cdot 10^{-11}$ between 1 and 18 $\lambda / D$. 
   }
   {
   The PESCC is a powerful, new focal-plane wavefront sensor that can be relatively easily integrated into existing ground-based {and future space-based} high-contrast imaging instruments. 
   }


\maketitle
\section{Introduction}\label{sec:introduction}
\begin{figure*}[!htb]
\centering
   \includegraphics[width=17cm]{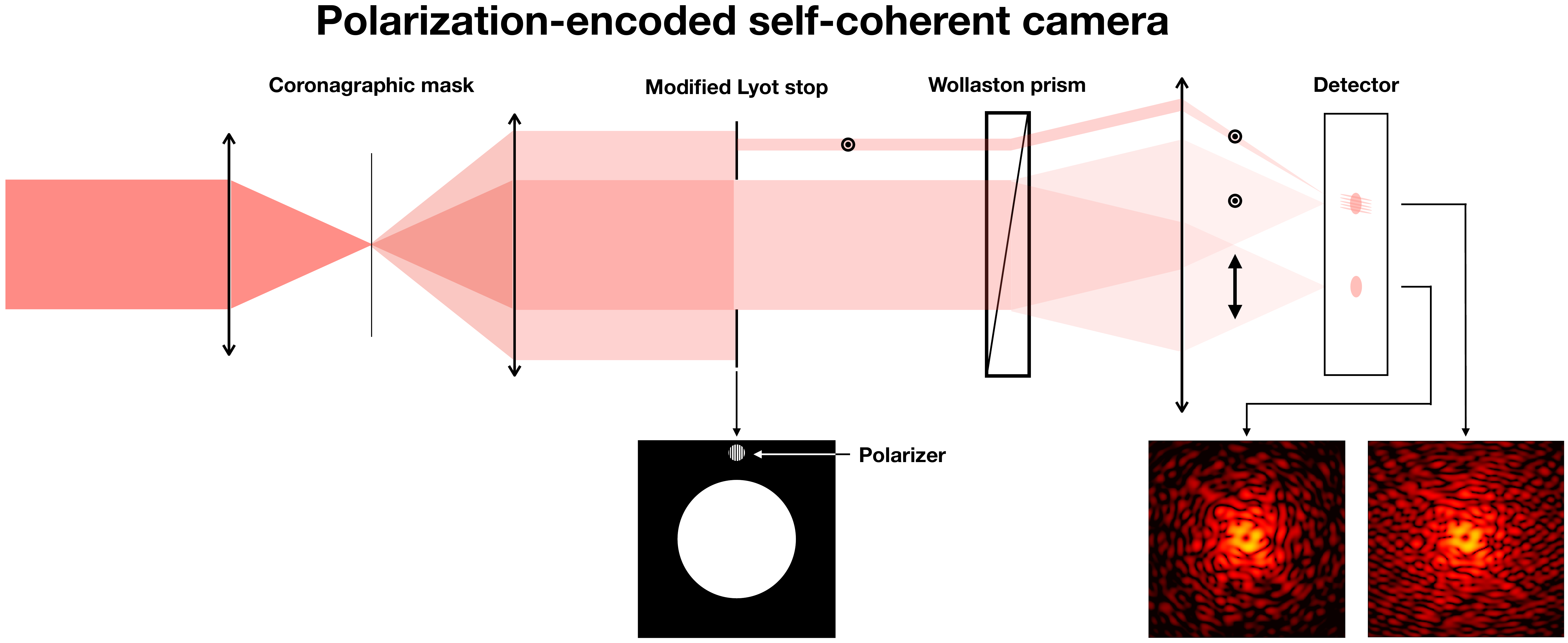}
     \caption{
     Overview of the system architecture of the PESCC. A focal-plane coronagraph (in this example a charge two vortex mask) diffracts the starlight out of the geometric pupil. 
     A reference hole (RH) with polarizer is placed in the Lyot stop and lets a polarized reference beam through. 
     The polarization state of the reference beam is pointing away from the page.   
     Prior to the focal-plane, a Wollaston prism (or any other polarizing beamsplitter) splits the two orthogonal polarization states. 
     The image with a polarization state that matches the RH polarizer has fringes that encode the wavefront information, while the orthogonal polarization state remains unmodulated by fringes.  
     }
    \label{fig:pol_scc_explanation}
\end{figure*} 
The direct imaging of exoplanets is a rapidly growing research field as it offers exciting opportunities in comparison to indirect methods such as the transit and radial velocity method. 
The star and exoplanet {light} are spatially separated and this therefore allows for the direct characterization of the exoplanet and (eventually) the search for biomarkers.
Furthermore, it can target exoplanets at wider separations than practically attainable with indirect methods and does not suffer from diminishing sensitivity when the exoplanet's orbit is not edge-on. 
However, the direct imaging of exoplanets is not a trivial task and many technical challenges still have to be solved.
For example, when observing an Earth-like exoplanet around a Solar-type star at 10 pc, the angular separation is expected to be $\sim$100 milliarcseconds and the contrast $\sim10^{10}$ in the visible (0.3 - 1 $\mu$m; \citealt{traub2010direct}). \\ 

\indent Modern ground-based high-contrast imaging (HCI) instruments (VLT/SPHERE \citealt{beuzit2019sphere}; Subaru/SCExAO \citealt{jovanovic2015subaru}; Gemini/GPI \citealt{macintosh2014first}; Magellan Clay/MagAO-X \citealt{males2018magao}, \citealt{close2018optical}) deploy extreme adaptive optics (XAO) systems to correct for wavefront errors caused by the turbulent Earth's atmosphere, coronagraphs to suppress starlight, and additional imaging, spectroscopic, and polarimetric post-processing techniques to further remove the speckle background. 
The current suite of observing and post-processing techniques consists of: angular differential imaging (ADI; \citealt{marois2006angular}), reference star differential imaging (RDI; \citealt{ruane2019reference}), spectral differential imaging (SDI; \citealt{sparks2002imaging}), polarimetric differential imaging (PDI; \citealt{kuhn2001imaging}), and coherent differential imaging (CDI; \citealt{guyon2004imaging}).
Both ADI and RDI are observing techniques that are relatively easy to implement, but have been suffering from temporal stability issues that prevent them to reach high contrasts at small separations.  
In addition, SDI has been shown to be ineffective at smaller inner working angles, as the radial movement of speckles by spectral diversity is minimal at these separations. However, PDI has achieved impressive results \citep{van2017combining} and can be simultaneously used for exoplanet characterization; still, exoplanets are never 100\% polarized, making this technique not the most efficient discovery method. 
With CDI, the light in the image is separated into its coherent and incoherent parts. 
As the exoplanet's light is by definition completely incoherent {with} the surrounding starlight, CDI can use 100\% of the exoplanet's light for the detection. 
This makes it a very promising technique, but it has thus far seen limited on-sky tests \citep{bottom2017speckle}. \\

\indent Typically, the XAO system consists of a deformable mirror (DM) with a high actuator count and a sensitive wavefront sensor, such as the Shack-Hartmann or Pyramid wavefront sensor, and delivers a high Strehl point-spread-function (PSF) to the instrument.
There is a non-common optical path difference between the science camera and the wavefront sensor split-off, in which aberrations occur due to misalignments and manufacturing errors. 
These so-called non-common path aberrations (NCPA) are not sensed by the main wavefront sensor and therefore left uncorrected. 
Due to changing temperature, humidity and gravity vector during observations, the NCPA slowly evolve, making them difficult to calibrate in post-processing and one of the current limitations in high-contrast imaging (\citealt{martinez2012speckle}; \citealt{martinez2013speckle}; \citealt{milli2016speckle}). 
One solution is to apply a focal-plane wavefront sensor (FPWFS) that uses the science images to measure the wavefront aberrations, which can subsequently be corrected by the DM. 
Ideally, the FPWFS is integrated with the coronagraph to enable simultaneous wavefront measurements and scientific observations.  
A host of different FPWFSs have been developed and tested on-sky \citep{jovanovic2018review}, but only a subset has a 100\% science duty cycle 
(\citep{codona2013focal, wilby2017coronagraphic, huby2017sky, guyon2017spectral, bos2019focal, miller2019spatial}). \\

\indent The FPWFS that is most relevant to this work is the self-coherent camera (SCC; \citealt{baudoz2005self}). 
The SCC places a reference hole (RH) in the Lyot stop of a coronagraph in an off-axis location. 
The RH transmits light that is diffracted by the coronagraph outside of the geometric pupil that would have otherwise been blocked by the Lyot stop. 
This light will propagate to the focal-plane, interfere with the on-axis beam, and generate high-spatial frequency fringes. 
The focal plane's electric field is spatially modulated and directly available by calculating the Fourier transform (FT) of the image.
This operation results in the optical transfer function (OTF) of the image.
The OTF of the SCC consists of three components: the central peak that is the PSF, and two sidebands, which are generated by the cross-talk between the fringes and PSF and which contain the electric field estimate. 
However, for the FT of the image to properly show the electric field estimate, the RH needs to be at least 1.5 times the pupil diameter from the center of the pupil.
This requires the system hosting the SCC to have large-diameter optics to contain both the reference and the central beam, and for the detector to have a high pixel density to properly sample the fringes in the focal plane.
The technique has been extensively tested in simulations \citep{galicher2010self} and in the lab \citep{mazoyer2013estimation, mazoyer2014high}; in the latter case, it has reached contrast levels of $\sim 5 \cdot 10^{-9}$ in narrow spectral bands \citep{potier2020comparing}. 
The SCC also enables CDI as post-processing technique \citep{galicher2007expected}, but requires additional flux measurements of the RH beam to calibrate the brightness of the RH PSF. 
It was shown in the lab \citep{singh2019active} and on sky \citep{galicher2019minimization} that the SCC is able to increase the contrast to by correcting the (quasi-)static aberrations during long exposure images with residual wavefront errors from the XAO system.
The major disadvantage of the SCC is the minimum distance of the RH at 1.5 times the pupil diameter, making it: 1) difficult to implement in existing instruments, as their optics do not have the required size; and 2) there is a little amount of light left this far from the on-axis beam, resulting in long exposure times for obtaining sufficient sensitivity. \\

\indent The Fast Atmospheric Self-coherent Camera Technique (FAST; \citealt{gerard2018fast}) tackles the second problem by modifying the focal-plane mask of the coronagraph to inject specifically more light into the RH. 
This provides the sensitivity {to} FAST for running at much shorter exposure times and {enables it to correct for} the rapidly changing residual wavefront errors from the XAO system. 
This technique was tested in simulation with a Lyot coronagraph, and more recently, with a vortex coronagraph \citep{gerard2020focal}.   
A lab test validated the coronagraph designs for the FAST concept \citep{gerard2019fast}.
Another recently developed concept is the fast-modulated SCC (FMSCC; \citealt{martinez2019fast}), which aims to address the other major disadvantage of the SCC. 
The FMSCC places the RH right next to the pupil, breaking the original minimum distance of 1.5 pupil diameters. 
In order to separate the sidebands from the central peak, which are now overlapping, in the OTF, two images are taken in quick succession with the RH blocked in one of the images.
The OTF of this second image only contains the central peak, and when it is subtracted from the first image, the two sidebands are revealed. 
For the subtraction to successfully reveal the sidebands on ground-based systems with the rapidly changing atmosphere, the system needs to either block the RH at high temporal frequency ($\sim$kHz) to "freeze" the atmosphere or to split the post-coronagraphic light into two beams:\ one with and one without RH. 
The first solution can only work for bright targets due to the high switching speeds, and the second solution is prone to differential aberrations between the two beams.
This concept bears great similarities with the differential OTF WFS (dOTF; \citealt{codona2013differential}), which extracts wavefront information by subtracting the OTFs of two PSFs, one of which is formed by an aperture with a small amplitude asymmetry.  \\  

\begin{figure}
\centering
   \includegraphics[width=\hsize]{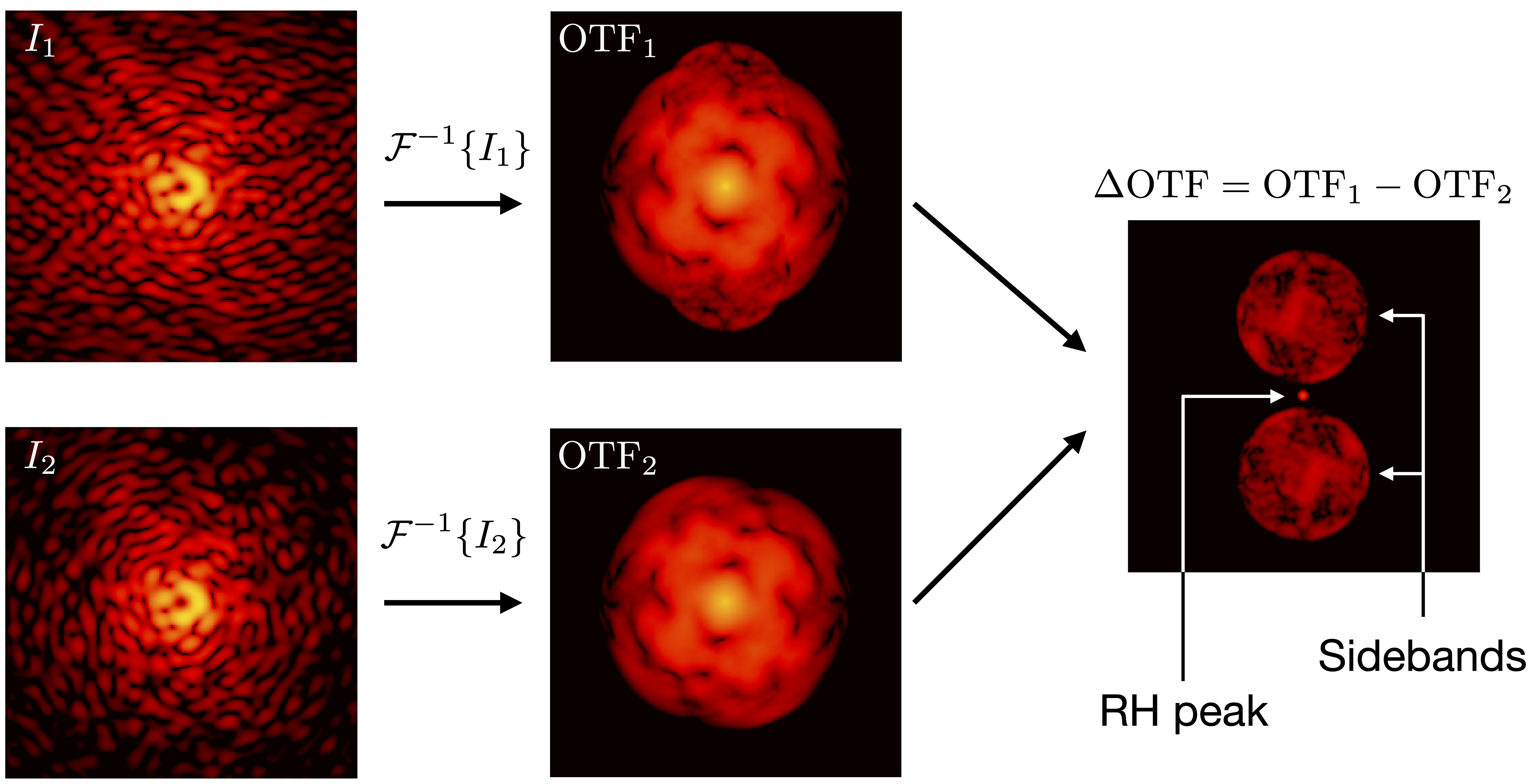}
     \caption{Example of sideband extraction with the PESCC.
                 For the images of channels 1 and 2 ($I_1$ and $I_2$) in \autoref{fig:pol_scc_explanation}, the OTFs are calculated. 
                 The OTF$_{1}$ contains the two sidebands with wavefront information and the RH peak, but these are overwhelmed by the central peak. 
                 When OTF$_2$, which only contains the central peak, is subtracted from OTF$_{1}$, the two sidebands and RH peak are revealed.  
     }
    \label{fig:OTF_example}
\end{figure}

\indent Here, we present the polarization-encoded self-coherent camera (PESCC), which is a variant of the FMSCC.
The PESCC features a polarizer in the RH that generates a polarized reference beam, which, in turn, generates fringes in one polarization state, while the orthogonal polarization state is unmodulated by fringes.
This is a concept that is very similar to the polarization differential OTF wavefront sensor \citep{brooks2016polarization}.
When the beam is split into two channels by a Wollaston prism just before the science camera, it ensures that there are minimal differential aberrations between the two polarization states.
See \autoref{fig:pol_scc_explanation} for an overview of the system architecture. 
The two polarization states can be imaged simultaneously, allowing for longer integration times and, therefore, fainter stars can be used as targets.   
For the images of the two polarization states, the OTF can then be calculated and subtracted to reveal the aberrated electric field (similar to the FMSCC analysis).
This process is shown in \autoref{fig:OTF_example}. 
Furthermore, as we show in \autoref{sec:theory}, the measurements also contain direct measurements of the RH, ensuring that the CDI is possible for every observation without additional measurements. 
We also show that because the RH is placed closer to the pupil, the PESCC relaxes the requirements on focal-plane sampling and this allows it to operate over broader wavelength ranges. 
Another advantage is that one of the two channels does not contain the reference beam and, therefore, it is not polluted by extra photon noise from the reference PSF. \\ 

\indent In \autoref{sec:theory}, we present the theory of the PESCC, including the CDI with the PESCC. 
We also carry out an analytical study of the performance of the PESCC compared to the SCC. 
In \autoref{sec:simulations}, we present the simulation results, specifically the wavefront sensing in \autoref{subsec:wavefront_sensing_sims}, wavefront control in \autoref{subsec:wavefront_control_sims}, and CDI in \autoref{subsec:CDI_sims}.
In \autoref{sec:conclusions}, we discuss the results and present our conclusions. 
\section{Theory}\label{sec:theory} 
\begin{table*}
\caption{Variables presented in \autoref{sec:theory}.}
\label{tab:theory_variables}
\vspace{2.5mm}
\centering
\begin{tabular}{l|l}
\hline
\hline
Variable & Description \\ \hline
$\beta$ & Factor that builds in safety margins in $\epsilon_0$. \\
$\gamma$ & Ratio of the pupil diameter and reference hole diameter. \\ 
$\delta \theta$ & Misalignment angle of the RH polarizer. \\ 
$\epsilon_0$ & Distance of the RH to the center of the pupil. \\ 
$\lambda$ & The wavelength of light. \\ 
$\Delta \lambda$ & The spectral bandwidth. \\
$d_r$ & Diameter of the reference hole. \\
$p$ & The strength of the instrumental polarization. \\
$t$ & The relative polarization leakage. \\
$D$ & Diameter of the pupil. \\
$E_{p}$ & The central beam pupil-plane electric field after the modified Lyot stop.\\ 
$E_{ref}$ & The pupil-plane electric field of the RH. \\
$\mathcal{F}\{\cdot\}$ & The Fourier transform operator. \\
$I_0$ & The focal-plane intensity of the central peak in the OTF. \\
$I_{ic}$ & The focal-plane intensity of the incoherent contribution (e.g., an exoplanet). \\ 
$I_{ct}$ & The focal-plane intensity of both sidebands in the OTF.  \\
$I_i$ & Focal-plane intensity image of channel $i$. \\
$I_i'$ & Focal-plane intensity image of channel $i$ with companion. \\
$I_{ref}$ & The focal-plane intensity of the RH. \\
$I_{ref}'$ & The focal-plane intensity of the RH corrected for polarizer leakage. \\
$I_{sb}$ & The focal-plane intensity of one sideband in the OTF.\\
$N_{act}$ & Number of actuators along one axis in the pupil. \\
$N_{pix}$ & Number of pixels per $\lambda/D$. \\ 
$\text{OTF}_i$ & The OTF of channel $i$. \\
$R_{\lambda}$ & The spectral resolution. \\
$\mathcal{S}$ & The detector sampling in units of pixels per $\lambda/D$. \\
\hline
\end{tabular}
\end{table*}

In this section, we focus on the theory behind the PESCC and perform an analytical study of its performance. 
We first derive the necessary equations in \autoref{subsec:PESCC}.
In \autoref{subsec:reference_hole_diameter} and \autoref{subsec:ref_hole_dist} we present the maximal RH diameter and its minimal distance from the pupil center. 
Using the equations derived in \autoref{subsec:ref_hole_dist}, we study how the smaller RH distance effects the focal-plane sampling constraints (\autoref{subsec:focal_sampling_constraints}) and the spectral bandwidth limitations (\autoref{subsec:bandwidth_limitations}). 
Then we investigate the effects of instrumental polarization and polarizer leakage in \autoref{subsec:instrumental_polarization} and \autoref{subsec:polarizer_leakage}.
Subsequently, we develop CDI with the PESCC in \autoref{subsec:CDI}. 
The variables presented in this section are defined in \autoref{tab:theory_variables}.
\subsection{Polarization-encoded self-coherent camera}\label{subsec:PESCC}
Here, we derive the working principle of the PESCC. 
We adopt the setup as shown in \autoref{fig:pol_scc_explanation}. 
The starlight first encounters a focal-plane coronagraph that diffracts it outside of the geometric pupil. 
The subsequent modified Lyot stop blocks most of the light and the RH, which contains a polarizer, transmits a fully polarized reference beam with an electric field, $E_{ref}$, with a diameter, $d_r$. 
The electric field of the central beam, directly after the Lyot stop, is given by $E_{p}$ and has a diameter, $D$. 
A polarizing beamsplitter (PBS) splits the beam into two channels that have orthogonal linear polarization states. 
The polarizer in the RH transmits a polarization state that is parallel to the polarization state of one of the channels. 
One of channels contains the reference beam and feature fringes in the focal-plane, the other does not. 
For now, we assume that the starlight is unpolarized, and that the polarizer featured in the RH and the polarizing beamsplitter (PBS) are perfect (i.e., they  split the light perfectly into two orthogonal polarization states and do not introduce wavefront aberrations).
In later subsections, we analytically and numerically investigate the consequences when these assumption do not hold. 
For simplicity, but without loss of generality, we also assume monochromatic, one-dimensional electric fields. 
The focal-plane intensities of channel 1 ($I_1$; with reference beam) and channel 2 ($I_2$; without reference beam), are given by:
\begin{align}
I_1 &= I_0 + I_{ref} + I_{ct}  \label{eq:intensity_1}, \\
I_2 &= I_0 \label{eq:intensity_2}, 
\end{align}
with $I_0 = |\mathcal{F} \{ E_{p}\}|^2$ as the PSF of the coronagraphic system, $I_{ref} = |\mathcal{F}\{E_{ref}\}|^2$ as the PSF of the RH, $I_{ct} = 2 \mathbb{R}\{ \mathcal{F}\{E_{p}\} \mathcal{F}\{E_{ref}\}^* \}$ as the cross-talk term between the RH and the central beam, and $\mathcal{F} \{ \cdot \}$ as the Fourier transform. 
{Here, $E^*$ denotes the complex conjugate of $E$.} 
The wavefront information of $E_{p}$ is encoded in $I_{ct}$, but is hidden behind the much stronger $I_0$ term. 
To retrieve the wavefront, we calculate the OTF of $I_1$ and $I_2$ with the inverse Fourier transform: 
\begin{align}
\text{OTF}_1 &= \mathcal{F}^{-1} \{ I_{1} \}, \\
                     &= E_{p} * E_{p}^* +  E_{ref} * E_{ref}^* +  \\
                     & \ \ \ \ \ E_{p} * E_{ref}^* * \delta(x+\epsilon_0)+ E_{p}^* * E_{ref} * \delta(x-\epsilon_0), \nonumber \\
\text{OTF}_2 &= \mathcal{F}^{-1} \{ I_{2} \}, \\
                    &= E_{p} * E_{p}^*, 
\end{align}
{with $*$ the convolution operator.} 
The OTF$_1$ consists of four terms, which are convolution combinations of the pupil-plane electric fields $E_{p}$ and $E_{ref}$. 
The term $E_{p} * E_{p}^*$ is the central peak in the OTF and is generated by the main beam in the system, its width in the OTF is $2D$.
The RH ($E_{ref} * E_{ref}^*$) also creates a peak at the same location as the main beam, but is much fainter than the central peak, and is smaller with a width of $2d_r$. 
The cross-talk between the RH beam and the central beam generates two lateral peaks or sidebands in the OTF located at $\pm \epsilon_0$, both with a width of $D+d_r$.
As the RH is placed close to the pupil ($\epsilon_0 < D$), the two sidebands still partly overlap with the central peak.  
See \autoref{fig:OTF_example} for a two-dimensional example that visualizes this. 
To reveal the sidebands, which contain the wavefront information, we subtract the $\text{OTF}_2$ from $\text{OTF}_1$:
\begin{align}\label{eq:delta_OTF}
\Delta \text{OTF} &= E_{ref} * E_{ref}^* + E_{p} * E_{ref}^* * \delta(x+\epsilon_0) + \\ 
                           & \ \ \ \  E_{p}^* * E_{ref} * \delta(x-\epsilon_0). \nonumber
\end{align}
If $\epsilon_0$ has been appropriately chosen (\autoref{subsec:ref_hole_dist}), the three remaining peaks in the OTF are well separated. 
It is essential that differential aberrations between the two channels are minimal, otherwise the $E_{p} * E_{p}^*$ term does not completely cancel in the subtraction. 
Extracting, centering and Fourier transforming one of the sidebands gives an estimation of the focal-plane speckle field $I_{sb}$ \citep{mazoyer2014high}: 
\begin{align}
I_{sb} &= \mathcal{F} \{ E_{p} * E_{ref}^* \} \label{eq:I_sb_estimation}, \\ 
          &= \mathcal{F} \{ E_{p} \}  \mathcal{F} \{ E_{ref}^* \}. 
\end{align} 
This term is used with wavefront sensing and control, as shown in \autoref{subsec:wavefront_control_sims}.
We note that $I_{sb}$ is a complex quantity, which becomes completely real when the other sideband is included. 
For CDI we also have to extract other information from the $\Delta \text{OTF}$. 
Selecting the RH peak and calculating its Fourier transform gives an estimate of the RH PSF: 
\begin{equation}\label{eq:I_ref_estimation}
I_{ref} = \mathcal{F} \{ E_{ref} * E_{ref}^* \} .
\end{equation}
It is also important to have an estimate of the cross-talk intensity term $I_{ct}$. 
This term can be estimated by extracting both sidebands and calculating their Fourier transform: 
\begin{align}
I_{ct} &= \mathcal{F}\{ E_{p} * E_{ref}^* * \delta(x+\epsilon_0) + E_{p}^* * E_{ref} * \delta(x-\epsilon_0) \}, \label{eq:I_ct_estimation} \\
        &= 2 \mathbb{R}\{ \mathcal{F}\{E_{p}\} \mathcal{F}\{E_{ref}'\}^* \}.
\end{align}

\subsection{Reference hole diameter}\label{subsec:reference_hole_diameter}
The diameter of the RH {($d_r$) directly} depends on the size of the dark hole. 
{The dark hole size is set by the maximum spatial frequency that can be controlled by the DM and is given by $\sqrt{2} N_{act} \lambda / (2D)$ \citep{mazoyer2013estimation}, with $D$ as the diameter of the pupil, $N_{act}$ the number of actuators along one axis in the pupil, and $\lambda$ the observed wavelength. } 
{The factor of $\sqrt{2}$ is included to account for the higher number of actuators along the diagonal compared to the sides of a square grid of actuators in a DM.}  
{To make sure that the DM can actually remove the speckles within the dark hole, the electric field of the speckles needs to be accurately measured.} 
{This can only happen when the focal-plane electric field of the RH is non-zero over the dark hole.} 
{The position of the first dark ring of the RH PSF is located at $1.22 \lambda / d_r$.}
Therefore, the diameter ($d_r$) is given by {\citep{mazoyer2014high}}:
\begin{equation}\label{eq:ref_hole_dia}
d_r \leq 1.22 \sqrt{2} \frac{D}{N_{act}}.
\end{equation}
Often, the ratio between the pupil diameter and reference hole diameter is used, $\gamma = D / d_r$. 
Then the \autoref{eq:ref_hole_dia} becomes:
\begin{equation}
\gamma \geq \frac{N_{act}}{1.22 \sqrt{2}}.
\end{equation}
\subsection{Reference hole distance}\label{subsec:ref_hole_dist}
\begin{figure}
\centering
\includegraphics[width=\hsize]{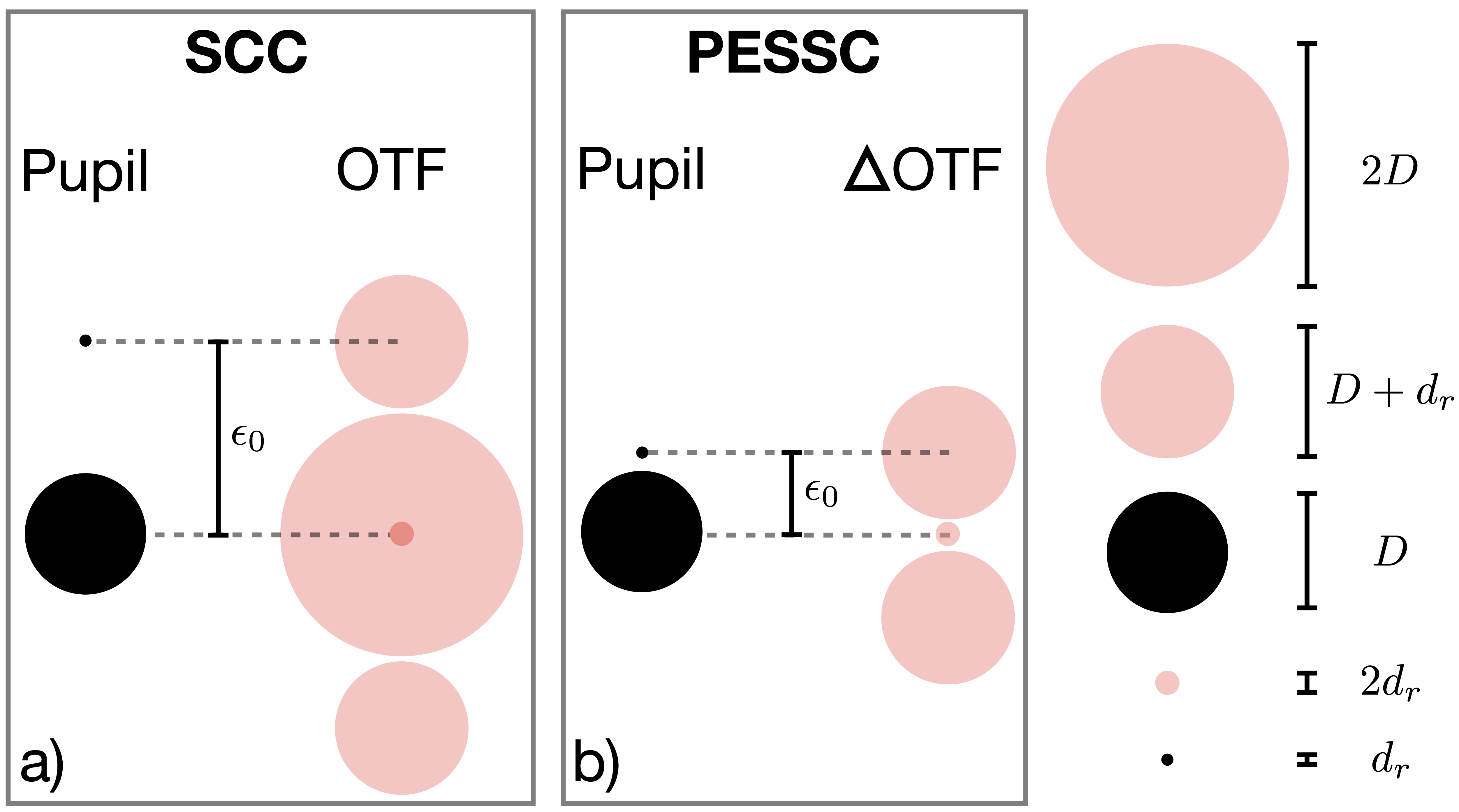}
\caption{
Explanation of how the geometry of the pupil relates to the geometry of the OTF. 
This figure shows that the minimum distance of the RH of (a) the SCC is much larger than that of (b) the PESCC. 
}
\label{fig:ref_hole_dist}
\end{figure}
For the SCC, the minimum distance of the reference hole ($\epsilon_0$) with respect to the center of the pupil was derived in \cite{galicher2010self}.
It ensures that the sidebands would not overlap with the central peak in the OTF (see \autoref{fig:ref_hole_dist} a), and is given by:
\begin{equation}\label{eq:ref_hole_dist_SCC}
\epsilon_0 = \frac{\beta}{2} \left(3 + \frac{1}{\gamma} \right) D,
\end{equation}
with $\beta$ a factor that cannot be lower than unity and usually set to 1.1 to include some extra margin. 
However, for the PESCC the sidebands can overlap with the central peak in the OTF as the central peak is subtracted out. 
The only constraint is that the sidebands do not overlap with one another or the peak from the RH, otherwise the wavefront information cannot be completely extracted, which is shown in \autoref{fig:ref_hole_dist} b. 
For the PESCC, this leads to the following reference hole distance law:
\begin{equation}\label{eq:ref_hole_dist_1_PESCC}
\epsilon_0 = \frac{\beta}{2} \left(1 + \frac{2}{\gamma} \right) D.
\end{equation}
Now we investigate in the ideal case how much closer the reference hole can be placed for the PESCC compared to the SCC. 
We set $\beta=1$ and assume an infinitely small reference hole $d_r \rightarrow 0$ ($\gamma \rightarrow \infty$). 
We then find for the SCC (\autoref{eq:ref_hole_dist_SCC}) that $\epsilon_0=1.5 D$. 
For the PESCC (\autoref{eq:ref_hole_dist_1_PESCC}) we find $\epsilon_0=0.5 D$. 
This means that the PESCC can be placed three times closer to the center of the pupil than the SCC.
This results in access to more light in the RH, as the focal-plane masks of coronagraphs diffracts more light closer to the geometric pupil, and, as shown in the following sections, it relaxes the focal-plane sampling constraints and allow for broader spectral bandwidths. 
\subsection{Focal-plane sampling constraints}\label{subsec:focal_sampling_constraints}
As the RH of the PESCC can be positioned significantly closer to the pupil, the constraints on the focal-plane sampling can be relaxed. 
In this subsection, we investigate what the sampling constrains are for the PESCC. 
For an unobstructed telescope pupil, the sampling ($\mathcal{S}$), in units of pixels ($N_{pix}$) per $\lambda/D$, is given by:  
\begin{equation}
\mathcal{S} = \frac{N_{pix}} {\lambda / D},
\end{equation}
with $\lambda$ as the wavelength. 
{To be Nyquist sampled} in the focal-plane, it is required that $N_{pix} \geq 2$, and is usually set to $N_{pix} = 3$ or $N_{pix} = 4$.
For the (PE)SCC, the combined diameter of the pupil and RH becomes $D \Rightarrow D + \epsilon_o - D/2 + d_r/2$ (\autoref{fig:ref_hole_dist}). 
This results in the following sampling constraint:
\begin{equation}
\mathcal{S} =  \frac{N_{pix}} {\lambda / [(D+d_r)/2 + \epsilon_0]}.
\end{equation}
When substituting the value of $\epsilon_0$ for the SCC (\autoref{eq:ref_hole_dist_SCC}) in this equation, we can rewrite it as: 
\begin{equation}\label{eq:sampling_SCC}
\mathcal{S} =  \frac{(1 + \beta)(1 + 1/ \gamma) + 2\beta}{2} \frac{N_{pix}}{\lambda / D}.
\end{equation}
For the PESCC (\autoref{eq:ref_hole_dist_1_PESCC}) we find: 
\begin{equation}\label{eq:sampling_PESCC}
\mathcal{S} = \frac{(1 + \beta)(1 + 1/ \gamma) + \beta / \gamma}{2} \frac{N_{pix}}{\lambda/D}.
\end{equation}
As in \autoref{subsec:ref_hole_dist}, we explore in an idealized example the gain in focal-plane sampling is for the PESCC.
Again, we assume that $\beta=1$, and that $d_r \rightarrow 0$ ($\gamma \rightarrow \infty$). 
We set $N_{pix}$ to 2 pixels to meet the Nyquist sampling constraint. 
For the SCC, we find that $\mathcal{S} = 4 $ pixels per $\lambda / D$, which means that the sampling should be twice as high compared to the unobscured pupil. 
On the other hand, the PESCC has a sampling of $\mathcal{S} = 2$ pixels per $\lambda / D$, which is equal to the case of the unobscured pupil. 
This is because in this idealized example, the RH is infinitely small and can be placed right on the edge of the pupil.  
This shows that the PESCC significantly relaxes the sampling requirements as the total number of pixels on the detector can be reduced, in the ideal case, by a factor of four.
\subsection{Spectral bandwidth limitations}\label{subsec:bandwidth_limitations}
\begin{figure}
\centering
\includegraphics[width=\hsize]{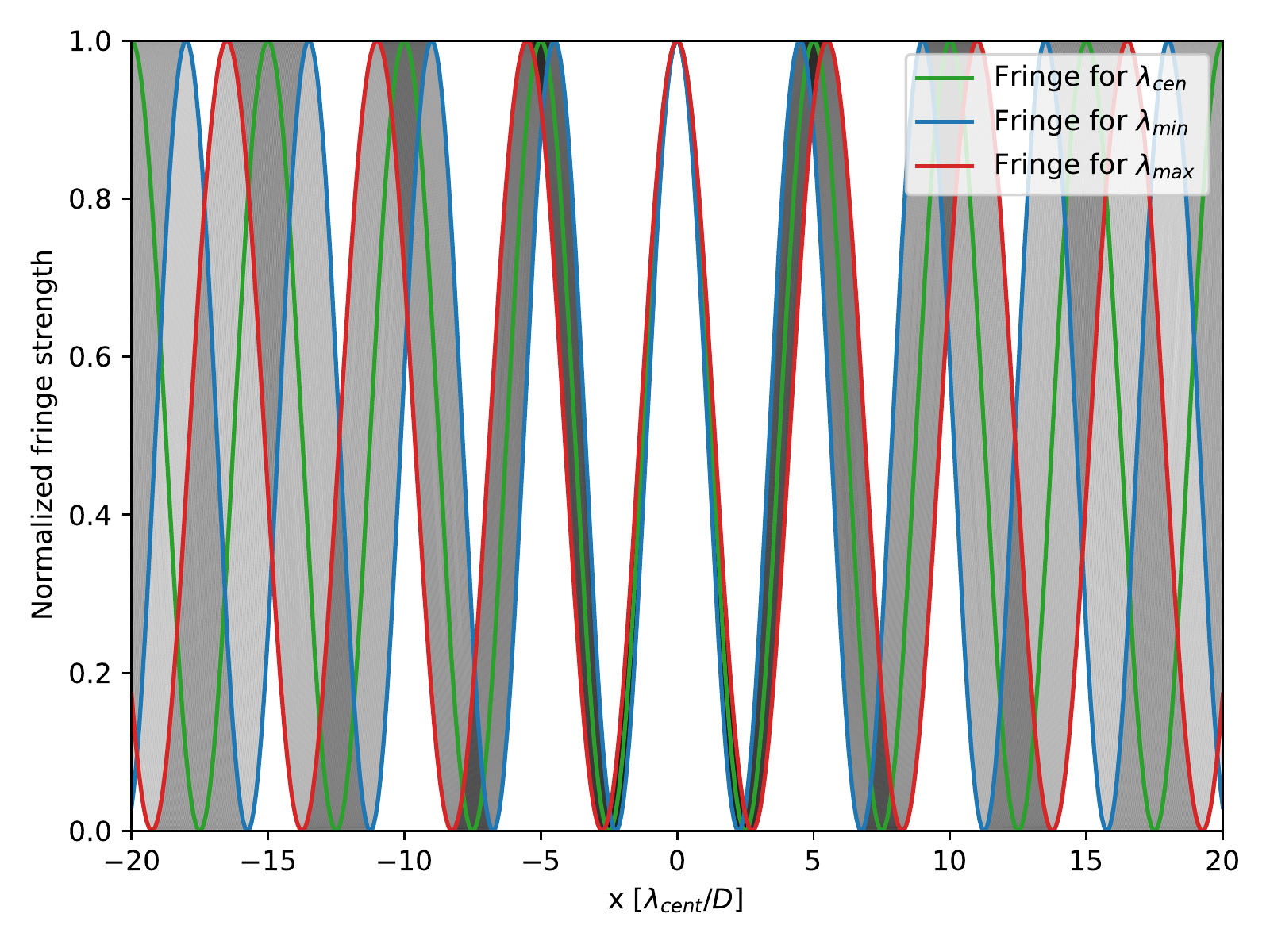}
\caption{
{Example of fringe smearing by spectral bandwidth. 
Fringes are plotted for many wavelengths within a broadband filter ($R_{\lambda} = 5$). 
All are plotted with a period of 5 $\lambda/D$, but due to the changing wavelength the physical fringe periods in the focal plane change as well. 
This results in significant fringe blurring after only a few periods from the center.}
}
\label{fig:fringe_smearing}
\end{figure}
Exoplanets are preferably observed over broad wavelength ranges ($\Delta \lambda$) to maximize the signal-to-noise ratio. 
However, the PSF is not constant, {changing its size with wavelength ($\propto \lambda/D$).} 
This means that the fringes introduced by {(PE)SCC also increase their size and period with wavelength.}
{Close to the center of the image these chromatic effects are not that pronounced, but after a few periods the fringes of the lowest and highest wavelengths in the filters start to significantly shift with respect to each other.} 
{An example of this is shown in \autoref{fig:fringe_smearing}.}
{After a certain distance from the image center the fringes are blurred to a level that it severely impacts the wavefront sensing performance.} 
{However, in the direction orthogonal to the fringe, the smearing is minimal and therefore the wavefront sensing in that direction is still relatively accurate. }
{Here, we study the bandwidth for (PE)SCC solutions with one RH. 
Broadband solutions for the SCC with multiple RHs do exist \citep{delorme2016focal}, but require even larger optics than the SCC to accommodate the additional reference beams.
This is because for SCC solutions with one DH, the constraint on the optics diameter can be somewhat mitigated by moving the central pupil from the center of the optics.
The RH beam and the main beam will then both pass through off-axis positions in the optics.  
However, for the broadband solutions with three RHs presented by \cite{delorme2016focal}, this is not the case because the RHs are evenly distributed around the central pupil such that it cannot be moved any more to reduce the optics diameter.}
\cite{galicher2010self} derived the minimal spectral resolution ($R_{\lambda} = \lambda_0 / \Delta \lambda$) required for {the} spectral smearing or chromatism not to impact the control region of the DM {(or dark hole size)}:
\begin{equation}
R_{\lambda} = \sqrt{2} N_{act} \frac{\epsilon_0}{D} - \frac{1}{2}.
\end{equation}
For the SCC (\autoref{eq:ref_hole_dist_SCC}), this becomes:
\begin{equation}\label{eq:spectral_resolution_SCC}
R_{\lambda} = \frac{\beta N_{act}}{\sqrt{2}} \left( 3 + \frac{1}{\gamma} \right) - \frac{1}{2}.
\end{equation} 
For the PESCC (\autoref{eq:ref_hole_dist_1_PESCC}), $R_{\lambda}$ is given by: 
\begin{equation}\label{eq:spectral_resolution_PESCC}
R_{\lambda} = \frac{\beta N_{act}}{\sqrt{2}} \left(1 + \frac{2}{\gamma} \right) - \frac{1}{2}.
\end{equation} 
We again go through our idealized example where we set $\beta=1$, $N_{act}=40$, and $d_r \rightarrow 0$ ($\gamma \rightarrow \infty$).
For the SCC, we find that $R_{\lambda} \approx 84$, which is consistent with the examples in \cite{galicher2010self}.
Then, for the PESCC, we find $R_{\lambda} \approx 24$.
Thus, the PESCC can operate over bandwidths that are $\sim$3.5 times wider than the SCC.
{This result can be understood as follows.}
{The period of the fringes is determined by $\epsilon_0$: when the RH is further away from the pupil, the period of fringes becomes shorter.}
{For a fixed spectral bandwidth the number of fringe periods before the blurring becomes too strong is also fixed.}
{Because the PESCC has a smaller $\epsilon_0$ than the $SCC$, the PESCC fringe periods are longer and the blurring becomes too strong at larger physical distances.}  
{Turning it around, when the size of the dark hole and thus the distance at which the blurring can occur are fixed, the PESCC can operate over broader bandwidths than the SCC.}
\subsection{Instrumental polarization}\label{subsec:instrumental_polarization}
\begin{figure}
\centering
   \includegraphics[width=\hsize]{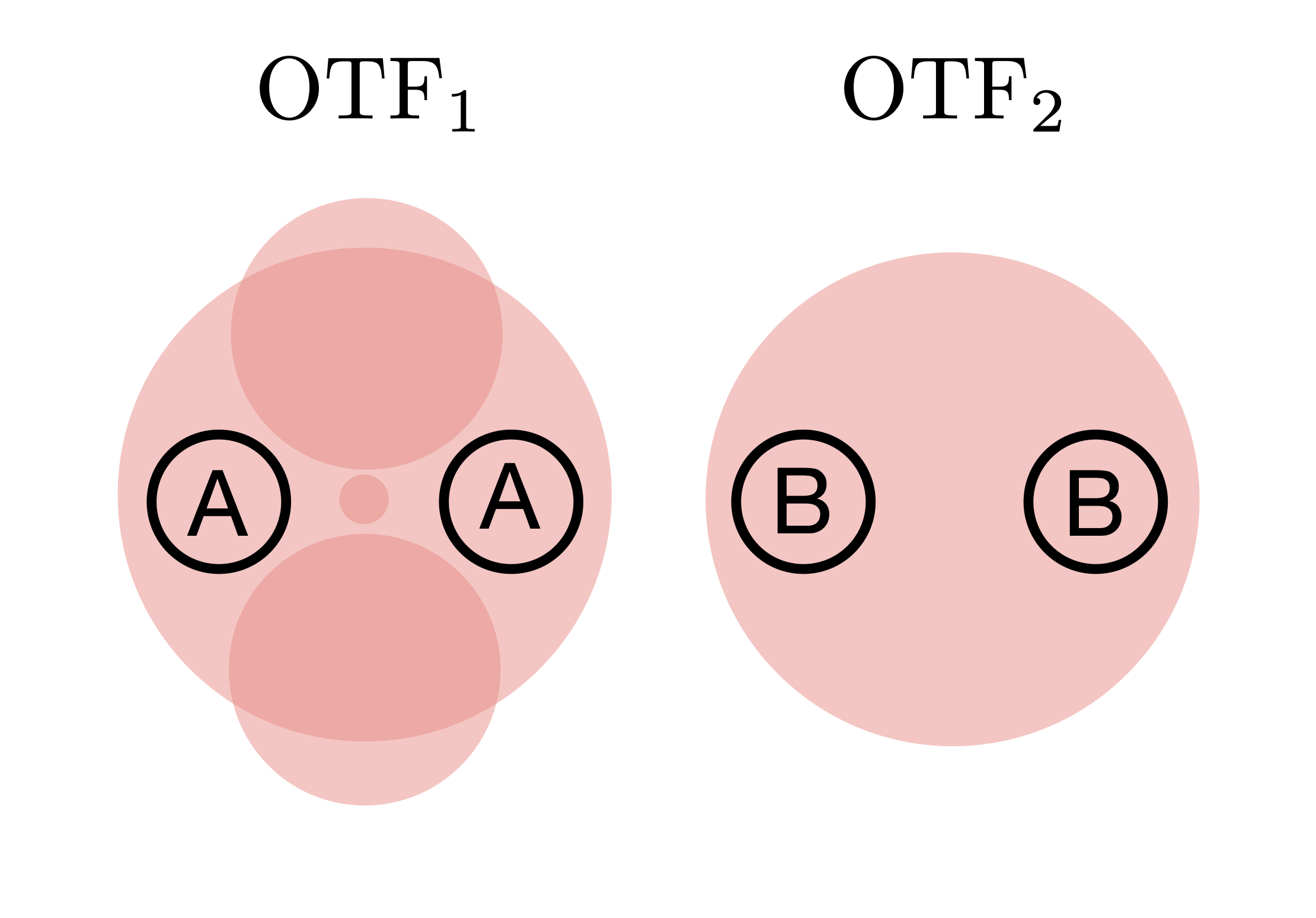}
     \caption{Regions in the $\text{OTF}_1$ and $\text{OTF}_1$ that are not contaminated by the sidebands or the RH. 
                Therefore, regions A and B are suitable to measure $p$.}
    \label{fig:OTF_dop_measurement}
\end{figure} 
When the starlight is polarized, it is possible that the performance of the PESCC is affected. 
This is because the central peaks in the two channels ($I_0$ in \autoref{eq:intensity_1} and \autoref{eq:intensity_2}) will not end up having an equal intensity and when the $\Delta \text{OTF}$ is calculated, they will not be completely canceled out in the subtraction. 
We are mainly concerned with the polarization introduced by the telescope and instrument because starlight is generally unpolarized (e.g., the integrated polarization signal of the Sun is $< 10^{-6}$; \citealt{kemp1987optical}). 
The instrumental polarization has been measured to be non-negligible for VLT/SPHERE and is on the level of $\sim10^{-2}$ \citep{van2020polarimetric}.
For the PESCC, we only have to account for the polarization states in which the PBS splits the light. 
We describe the instrumental polarization's strength by $p = (I_{0,1} - I_{0,2}) / (I_{0,1} + I_{0,2})$ ($-1 \leq p \leq 1$), with $I_{0,x}$ as the intensity of the central peak in channel, $x$. 
If $p=1$, then the light is polarized such that all the light is in channel 1; if $p=-1,$ all the light is in channel 2; and when $p=0,$ both channels have equal amounts of light. 
This can be viewed as a normalized Stokes parameter. 
Here, we explore the effect of a non-zero $p$.
We rewrite \autoref{eq:delta_OTF} such that it includes these polarization effects:  
\begin{align}
\Delta \text{OTF} &= p E_{p} * E_{p}^* + \frac{1 + p}{2} E_{ref} * E_{ref}^* + (1 + p) \cdot  \label{eq:delta_OTF_dop} \\ 
                           & \ \ \ \  [E_{p} * E_{ref}^* * \delta(x+\epsilon_0) + E_{p}^* * E_{ref} * \delta(x-\epsilon_0)].  \nonumber
\end{align}
The major problem at hand is the residual of the central peak ($p E_{p} * E_{p}^*$), as it will overlap with the sidebands and affect the wavefront estimate, particularly because the central peak is much brighter than the sidebands. 
The $1+p$ factor that affects the RH peak and the sidebands will put more photons in these terms when $p>0$ and fewer photons when $p<0$, and therefore it will affect the wavefront sensing sensitivity by decreasing or increasing the photon noise.
However, as $p$ is expected to be around $10^{-2}$, this effect will have less of an impact than the residuals of the central peak. \\
\indent If the value of $p$ is known, then it is possible to compensate for its effects in post-processing by dividing $\text{OTF}_1$ and $\text{OTF}_2$ with, respectively, $1+p$ and $1-p$. 
This removes the detrimental effects of the first term in \autoref{eq:delta_OTF_dop}, but it does not affect the sensitivity of the other terms as it cannot correct the fundamental effects of photon noise. 
Much effort has already gone in understanding the polarization effects of high-contrast imaging instruments (\citealt{de2020polarimetric}; \citealt{van2020polarimetric}).
Therefore, a detailed model that describes the instrumental polarization at a given configuration of the telescope and instrument could help to mitigate these effects.
It is also possible to directly measure $p$ in $\text{OTF}_1$ and $\text{OTF}_2$.
As shown in \autoref{fig:OTF_dop_measurement} by the circles A and B, it is possible to selection regions in the OTFs without contamination of the sidebands or the RH.
By calculating the flux in region A ($F_A$) and B ($F_B$), it is possible to estimate $p$: 
\begin{equation}\label{eq:IP_correction}
p = \frac{F_A - F_B}{F_A + F_B}.
\end{equation}
\subsection{Polarization leakage}\label{subsec:polarizer_leakage}
In \autoref{subsec:PESCC}, we assume that the RH polarizer and the PBS would perfectly split the two polarization states.
However, polarizers are not perfect and can be misaligned with respect to each other, which makes the reference beam leak from channel 1 into channel 2. 
Channel 2 will then also form (weaker) fringes in the focal-plane image, which results in the sidebands in the OTF. 
When calculating $\Delta$OTF, as in \autoref{eq:delta_OTF}, these sidebands in channel 2 will remove the signal from the sidebands in channel 1, affecting the wavefront estimates.  
We rewrite \autoref{eq:delta_OTF} such that it includes the polarization leakage: 
\begin{align}
\Delta \text{OTF} &= \frac{1-t}{2} E_{ref} * E_{ref}^* + (1-\sqrt{t}) \cdot \label{eq:delta_OTF_polarizer_leakage} \\ 
                           & \ \ \ \  [E_{p} * E_{ref}^* * \delta(x+\epsilon_0) + E_{p}^* * E_{ref} * \delta(x-\epsilon_0)],  \nonumber
\end{align}
with $t$  ($0 \leq t \leq 1$) the relative level of polarization leakage, which simulates the extinction ratio of the polarizer as $1 / t$, and the effect of a misaligned polarizer as: 
\begin{equation}\label{eq:misalignment_pol_leakage}
t = 2 \sin^2(\delta \theta),
\end{equation} 
with the misalignment angle $\delta \theta$.
\autoref{eq:delta_OTF_polarizer_leakage} shows that the central peak is always be subtracted out in this case, which is not surprising because the main effect of polarization leakage is the reference beam leaking into channel 2. 
It also shows that the accuracy of the sideband estimate is now proportional to $1-\sqrt{t}$, effectively reducing the response of the wavefront sensor, which will eventually lower the gain of the wavefront control loop. 
The accuracy with which the flux in the reference hole can be estimated scales more favorably with $(1-t)$ and is therefore less affected. 
When, for example, the RH polarizer has an extinction ratio of 100:1, that is, $ 1/t = 100 \rightarrow t=10^{-2}$, then the wavefront estimate is 90\% of its true value, while the reference hole flux is 99\% of the truth. 
If the extinction ratio is accurately known, for example by measurements before installing the modified Lyot mask, this effect can be corrected for during post-processing or wavefront control.
Another effect to take into account, especially with CDI as discussed in \autoref{subsec:CDI}, is that the polarization leakage also contaminates the second channel. 
This introduces an extra source of photon noise during post-processing with CDI.
Therefore, although the effects on the wavefront control can be calibrated, a high performance polarizer with low leakage is desirable. 
As discussed in \autoref{sec:conclusions}, a prime candidate for the RH polarizer are wire grids polarizers. 
These can be manufactured to have an extinction ratio of {1000:1 - 10.000:1 \citep{george2013improved}}, which would result in $t=10^{-3}-10^{-4}$.  
When the PBS is implemented as a Wollaston prism, which we foresee to be used, the extinction ratio exceeds 100.000:1 \citep{king1971some}, which is equivalent to $t\leq10^{-5}$ and thus negligible compared to the wire grid polarizer performance. 
A rotational misalignment between the RH polarizer and PBS gives a polarization leakage dictated by \autoref{eq:misalignment_pol_leakage}. 
For misalignments of 1$^{\circ}$, 3$^{\circ}$, and 5$^{\circ}$, we find $t\approx6\cdot10^{-4}$, $t\approx5\cdot10^{-3}$, and $t\approx2\cdot10^{-2}$, respectively. 
Therefore, it is likely that rotational misalignments will dominate the polarization leakage. 
\subsection{Coherent differential imaging}\label{subsec:CDI}
In this subsection, we investigate how CDI is performed with the PESCC. 
If we consider adding the light of an unpolarized, incoherent circumstellar environment (e.g., an exoplanet, or circumstellar disk), the measurements in channel 1 and 2 become: 
\begin{align}
I_1' &= I_1 + I_{ic}, \\
I_2' &= I_2 + I_{ic},
\end{align}
with $I_1$ and $I_2$ given by \autoref{eq:intensity_1} and \autoref{eq:intensity_2}, and $I_{ic}$ the incoherent contribution. 
We assume that $p=0$ and $t=0$.
Deriving the $I_c$ term is slightly different for the two channels, as only one has the RH beam interfering. 
For channel 1 we find $I_{ic}$ as \citep{galicher2007expected}:
\begin{equation}\label{eq:incoherent_companion_channel_1} 
I_{ic} = I_1' - I_{ref} - I_{sb}^2 / I_{ref} - I_{ct},
\end{equation}
with $I_{ref}$ given by \autoref{eq:I_ref_estimation}, $I_{sb}$ given by \autoref{eq:I_sb_estimation}, and $I_{ct}$ given by \autoref{eq:I_ct_estimation}.
For channel 2 we find $I_{ic}$ as:
\begin{equation}\label{eq:incoherent_companion_channel_2}  
I_{ic} = I_2' -  I_{sb} / I_{ref}.
\end{equation}
We note that the second channel does not contain the RH PSF, and is therefore not affected by photon noise from this term. 
These equations only hold for the perfect system. 
When there are system inaccuracies present, the CDI performance will degrade significantly. 
However, inaccuracies such as $p$ and $t$ can be accounted for in the post-processing step if the correct values are known. 
As shown in \autoref{subsec:instrumental_polarization}, it is possible to measure $p$ in the $\text{OTF}_1$ and $\text{OTF}_2$. 
For $t$, it would have to be measured preferably before the modified Lyot stop is installed.  
When correcting for these effects, \autoref{eq:incoherent_companion_channel_1} and \autoref{eq:incoherent_companion_channel_2} become: 
\begin{align}
I_{ic} &= I_1'' - I_{ref}' - I_{sb}'^2 / I_{ref}' - I_{ct}', \\
I_{ic} &= I_2'' - t^2 I_{ref}' - I_{sb}'^2 / I_{ref}' - t I_{ct}',
\end{align} 
with $I_1'' = I'_1 / (1+p)$ and $I_2'' = I'_2 / (1-p)$ the images corrected for instrumental polarization effects, $I_{ref}' = I_{ref} / (1 - t^2)$, $I_{sb}' = I_{sb} / (1 - t)$, and $I_{ct}' = I_{ct} / (1-t)$ the correction for polarizer leakage of respectively the RH beam, the intensity of the sideband, and the intensity of the cross-talk term. 
\section{Simulations}\label{sec:simulations}
\begin{table}
\caption{Simulation parameters for \autoref{sec:simulations}.}
\label{tab:simulation_parameters}
\vspace{2.5mm}
\centering
\begin{tabular}{ll}
\hline
\hline
Variable & Value \\ \hline
$\lambda$ & 1550 nm \\ 
Pupil diameter & 7 mm \\
Aperture & Clear \\ 
Coronagraph & Charge two vortex  \\ 
Lyot stop diameter & 0.99 pupil diameter\\ 
Deformable mirror & 40 $\times$ 40 actuators \\
\\
$\beta$ & 1.1 \\ 
$d_r$ & 0.3 mm \\
$\gamma$ & 23.2 \\
$\epsilon_0$ PESCC & 4.2 mm \\
$\epsilon_0$ SCC &  11.7 mm \\
\\
Focal-plane sampling PESCC & $2.24$ pixels per $\lambda/D$ \\
Focal-plane sampling SCC & $4.39$ pixels per $\lambda/D$ \\
\hline
\end{tabular}
\end{table}
In this section, we investigate the performance of the PESCC in numerical simulations, and compare it to the SCC where relevant. 
The effects of photon noise, differential aberrations, instrumental polarization, polarizer leakage, and spectral resolution on wavefront sensing and control are explored.  
The simulations are performed in Python using the {HCIPy package \citep{por2018hcipy}}, which includes polarization propagation with Jones matrices necessary for this work.
We simulate an idealized HCI system with static wavefront aberrations. 
The system operates at 1550 nm and consists of a clear aperture, {an idealized DM (e.g., no actuator cross-talk, or quantization errors) with a 40$\times$40 square grid of actuators located in the pupil plane}, a charge two vortex coronagraph, a (PE)SCC Lyot stop, a polarizing beamsplitter and detector. 
The diameter of the pupil before the vortex coronagraph is 7 mm.
The wavefront aberrations that are looked at are induced by an out-of-plane phase aberration {following a power spectral density with a power law exponent of -3.} 
Fresnel propagation from this plane to the pupil creates both phase and amplitude aberrations.
This results in a wavefront error (WFE) of $2.9\cdot10^{-2}$ $(\pm8\cdot10^{-3}$) $\lambda$ root mean square (RMS), and intensity variations over the pupil of $16$ $(\pm1$\%) RMS (measured over 100 random aberrations). 
The Lyot is undersized by 1\% compared to the pupil diameter.
The diameter of the RH is determined by \autoref{eq:ref_hole_dia} and for this specific system it is set to 0.3 mm ($\gamma=23.2$).
The RH distance is determined by \autoref{eq:ref_hole_dist_SCC} and \autoref{eq:ref_hole_dist_1_PESCC} for the SCC and PESCC respectively. 
For $\beta=1.1$, the RH distance is 11.7 mm for the SCC and 4.2 mm for the PESCC.
This automatically sets the focal-plane sampling to $4.39$ pixels per $\lambda/D$ for the SCC (\autoref{eq:sampling_SCC}) and $2.24$ pixels per $\lambda/D$ for the PESCC (\autoref{eq:sampling_PESCC}). 
The simulation parameters are summarized in \autoref{tab:simulation_parameters}. \\
\indent First, we investigate the wavefront sensing performance in \autoref{subsec:wavefront_sensing_sims}.
Subsequently, we look into wavefront sensing and control in \autoref{subsec:wavefront_control_sims}.
Finally, we test the CDI with the PESCC in an idealized system in \autoref{subsec:CDI_sims}. 
As the wavefront aberrations considered here remain static during the simulations, the conditions and results are more representative for space-based observatories. 
They mainly serve as proof of principle, and in a future work, we will investigate more realistic conditions for ground-based observatories.  
\subsection{Wavefront sensing}\label{subsec:wavefront_sensing_sims}
In this subsection, we investigate how the wavefront sensing capabilities of the PESCC compare to the SCC and how they degrade due to various noise sources. 
To estimate the wavefront sensing performance, we calculate, from the two channels images, the $\Delta \text{OTF}$ as in \autoref{eq:delta_OTF} when the noise source is applied. 
Subsequently, we select one of the sidebands and center it. 
This sideband is the pupil-plane electric field convolved with the pupil-plane RH electric field, and we consider that to be the pupil-plane electric field estimate. 
Similarly, a noiseless pupil-plane electric estimate is calculated for the same wavefront aberration. 
The residual RMS wavefront error, which is the WFE common to both channels, is calculated by subtracting the phase of the noiseless electric field estimate from the estimate with noise, and is converted from units of radians to relative units of fractional $\lambda$. 
We simulate various levels of every noise source, and for every level, we simulate a hundred random wavefront aberration instances. 
\subsubsection{Photon noise performance}
\begin{figure}
\centering
\includegraphics[width=\hsize]{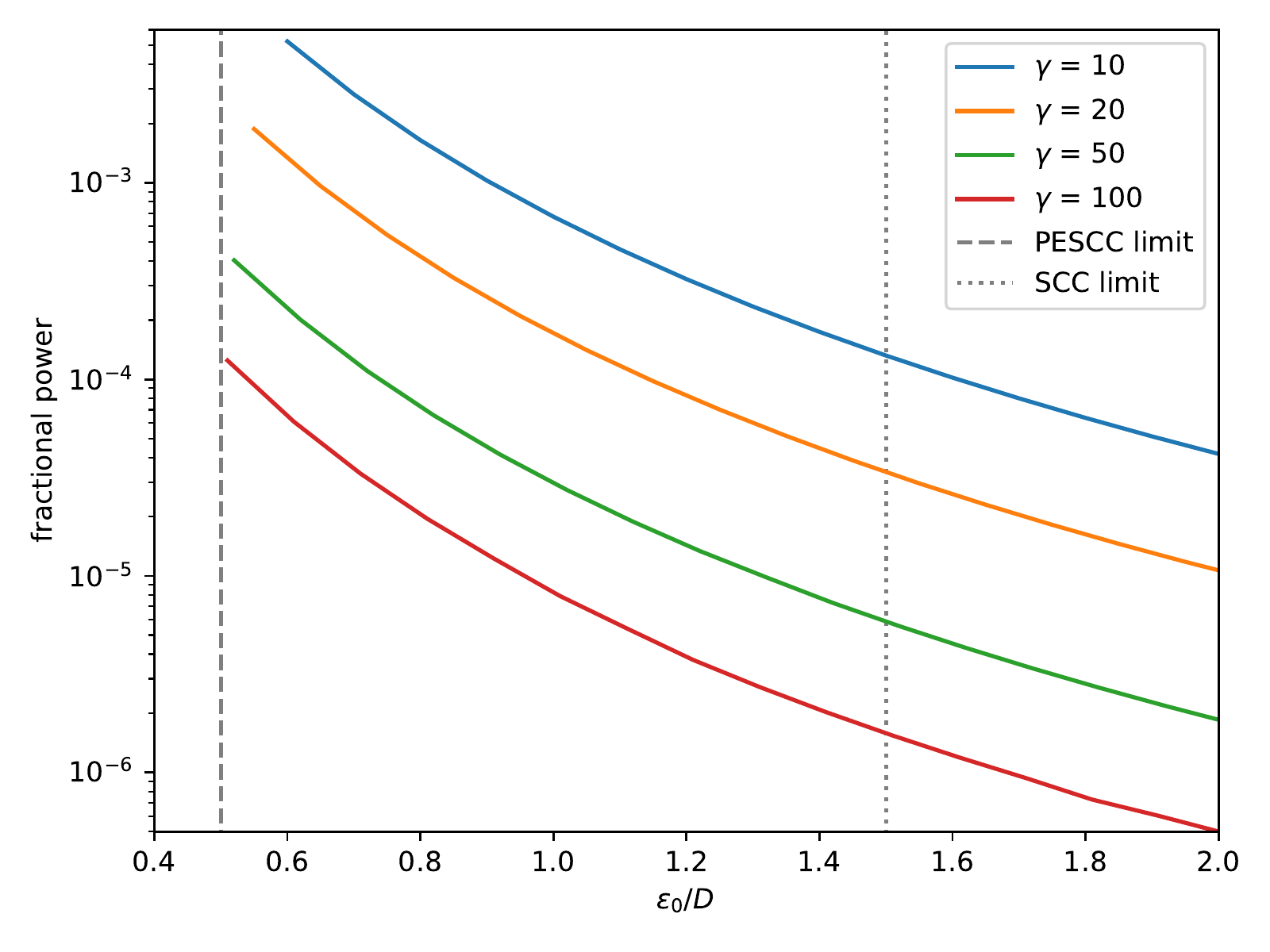}
\caption{Fractional power in the RH (ratio of the power in the RH and total power in pupil plane before Lyot stop) as function of distance from the center of the pupil.
             This was simulated using a charge two vortex coronagraph and does not include the polarizer in the RH for the PESCC to show the total power available. 
             }
\label{fig:fractional_power}
\end{figure}
\begin{figure}
\centering
\includegraphics[width=\hsize]{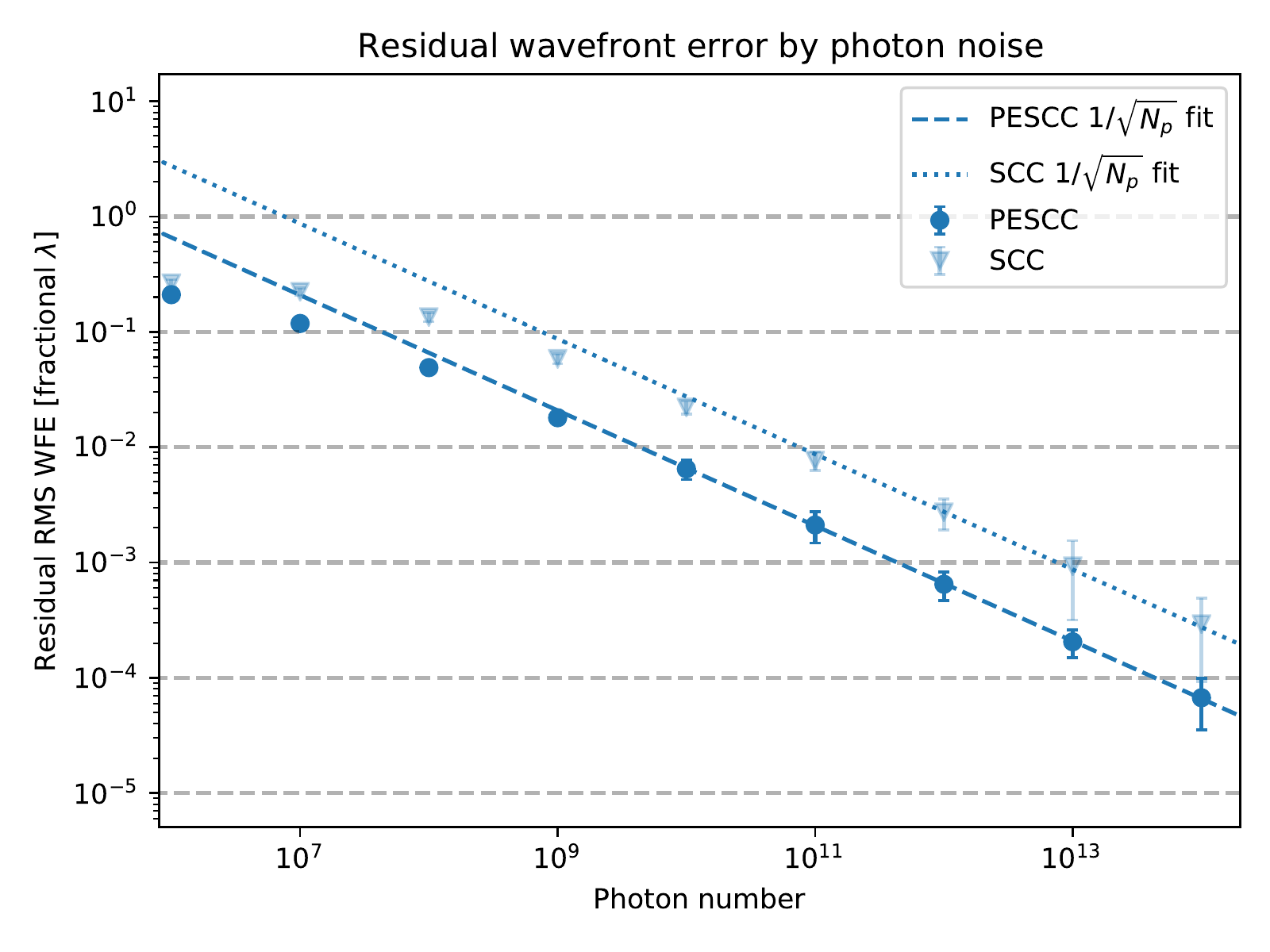}
\caption{
{Photon noise sensitivity of the PESCC and SCC for wavefront sensing. 
The reported photon number is the number of photons before the coronagraph. 
The error bars show the $1\sigma$ deviation over the 100 random wavefront aberration instances. 
The dashed and dotted lines show $1/\sqrt{N_p}$ fits for photon numbers $\geq10^{11}$ to show the regimes in which the performance is photon noise-limited. 
}
}
\label{fig:photon_noise_sensitivity}
\end{figure}
As discussed in \autoref{subsec:ref_hole_dist}, the RH of the PESCC can be positioned much closer to the pupil compared to the RH of the SCC. 
This provides access to a greater number of photons for wavefront sensing as the coronagraph's focal-plane mask scatters most light close to the geometric pupil. 
In \autoref{fig:fractional_power}, we plot the fractional power of the RH (ratio of the power in the RH and the total power available) as function of the distance to the pupil center ($\epsilon_0 / D$) for various $\gamma$ with a charge 2 vortex coronagraph. 
We do not include the polarizer that would be installed in the RH of the PESCC to show the total power available. 
This figure shows that there is a greater number of photons available for the PESCC compared to the SCC, for example, at their respective minimum $\epsilon_0$ (shown in the figure with the vertical, dotted and dashed lines), there is a factor of $\sim$64 difference (factor $\sim$32 including the polarizer). 
However, the polarizing beamsplitter does also split the main beam into two, which effectively halves the number of photons in the main beam. 
Therefore, the effective increase in the photon numbers is $\sim$16.  
We note that this last sensitivity hit only applies to wavefront sensing and not for companion detection because in the latter case, both channels can be combined. 
To indicate the expected wavefront sensing performance, we plot in \autoref{fig:photon_noise_sensitivity} the wavefront sensing performance as function of the number of photons before the coronagraph. 
This figure shows that the PESCC consistently outperforms the SCC by a factor of four, which is to be expected as the PESCC receives $\sim$16 more photons ($\sqrt{16}=4$). 
This enables the PESCC to either achieve a sensitivity that is four times higher or run with wavefront control loop speed that is a 16 times higher. 
The dotted and dashed lines show $1/\sqrt{N_p}$ fits ($N_p$ is the photon number) that are fitted to the data points for $N_p \geq 10^{11}$ photons. 
When the photon numbers become too low, then there are not enough photons in the sidebands for wavefront information to be subtracted and the noise becomes dominated by numerical artifacts, which explains the flattening of the data points.  
As the PESCC has access to more photons, it occurs at the lower photon number. 
For $10^{12}$ photons, the PESCC reaches a $<10^{-3}$ $\lambda$ RMS WFE, a similar WFE is reached by the SCC at $\sim 1.6 \cdot 10^{13}$ photons. 
 \subsubsection{Differential aberrations}\label{subsubsec:differential_aberrations_wfs}
\begin{figure}
\centering
\includegraphics[width=\hsize]{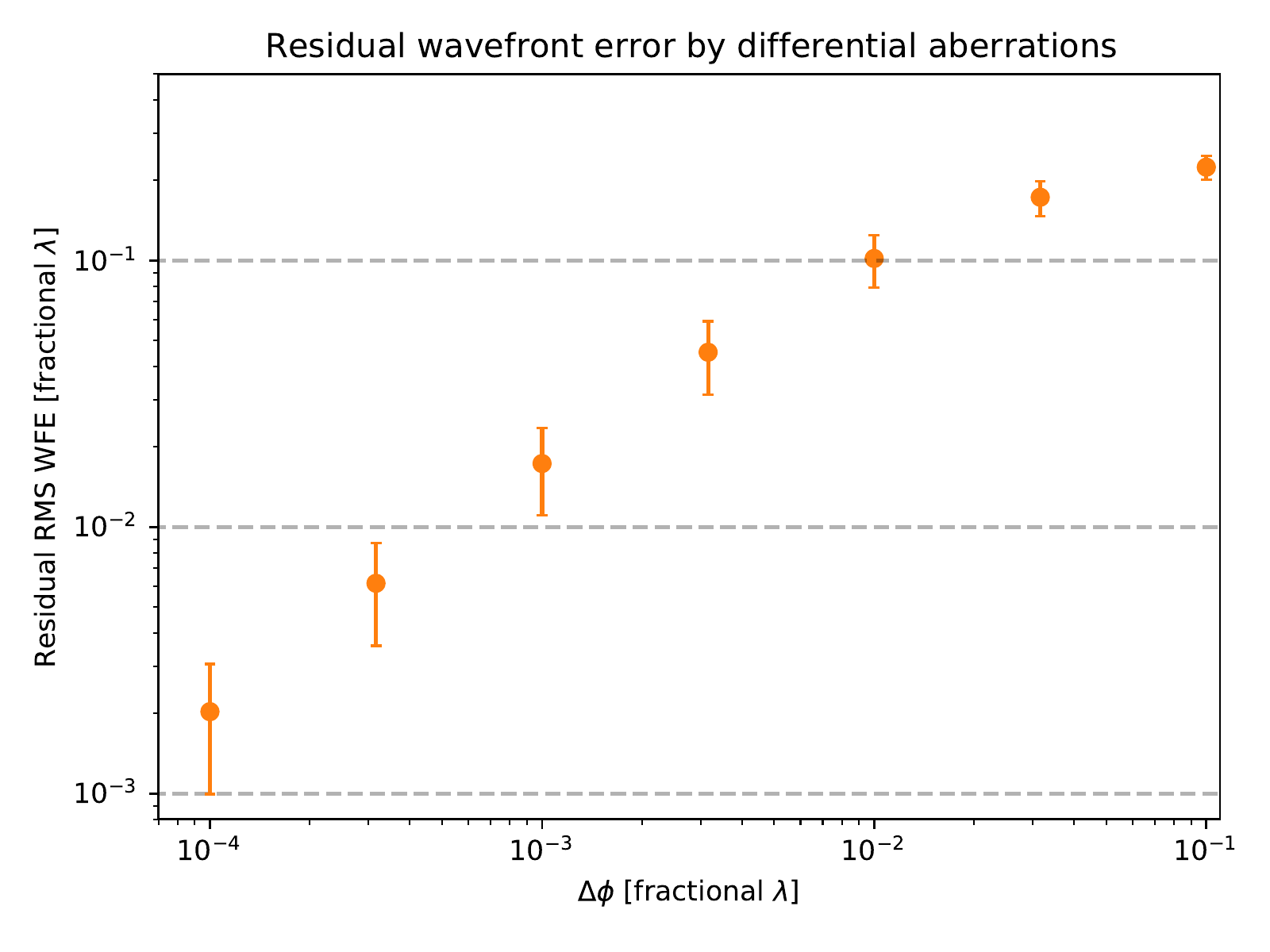}
\caption{
RMS differential aberrations ($\Delta \phi$) between the two beams downstream of the polarizing beamsplitter, and their effect on the wavefront reconstruction. 
The error bars show the $1\sigma$ deviation over the 100 random wavefront aberration instances. 
}
\label{fig:differential_aberrations}
\end{figure}
The principle of the PESCC is that the OTFs of the two beams can be subtracted to reveal the sidebands. 
However, these beams follow different optical paths in and downstream of the PBS.
A converging beam propagating through a PBS such as a Wollaston prism can incur differential aberrations between the outgoing beams \citep{simon1986wollaston} and these differential aberrations can increase when the beams hit the downstream at optics at slightly different positions. 
These differential aberrations introduce residuals in the $\Delta \text{OTF}$ that affect the wavefront estimation. 
Here, we quantify the effect of these differential aberrations on the wavefront estimation. 
We assume that that the polarizing beam splitter is one of the last elements in the optical train and that downstream optics only introduce low-order aberrations from misalignments. 
We simulate the differential aberrations by introducing on one beam a random combination of seven low order aberrations (starting at defocus), that have been scaled to a certain rms wavefront error ($\Delta \phi$). 
These low-order aberrations include differential aberrations expected by the Wollaston prism \citep{simon1986wollaston} and from the downstream optics (derived from Zemax simulations).
The other beam does not get any additional aberrations, so that we make sure that we tightly control the level of differential aberrations. 
In \autoref{fig:differential_aberrations}, we plot wavefront estimation performance as a function of the level of differential aberration. 
It shows that the wavefront is severely affected by differential aberrations. 
For $\Delta \phi \approx 5 \cdot 10^{-4}$ $\lambda$, the residual RMS WFE is $\sim10^{-2}$ $\lambda$.
This means that differential aberrations will have to be tightly controlled for successful operation of the PESCC. 
{To put these values into perspective, SPHERE/IRDIS was built with $\sim6 \cdot 10^{-3}$ waves of differential aberrations \citep{dohlen2008infra} between the two beams.}
 \subsubsection{Instrumental polarization}
  \begin{figure}
\centering
\includegraphics[width=\hsize]{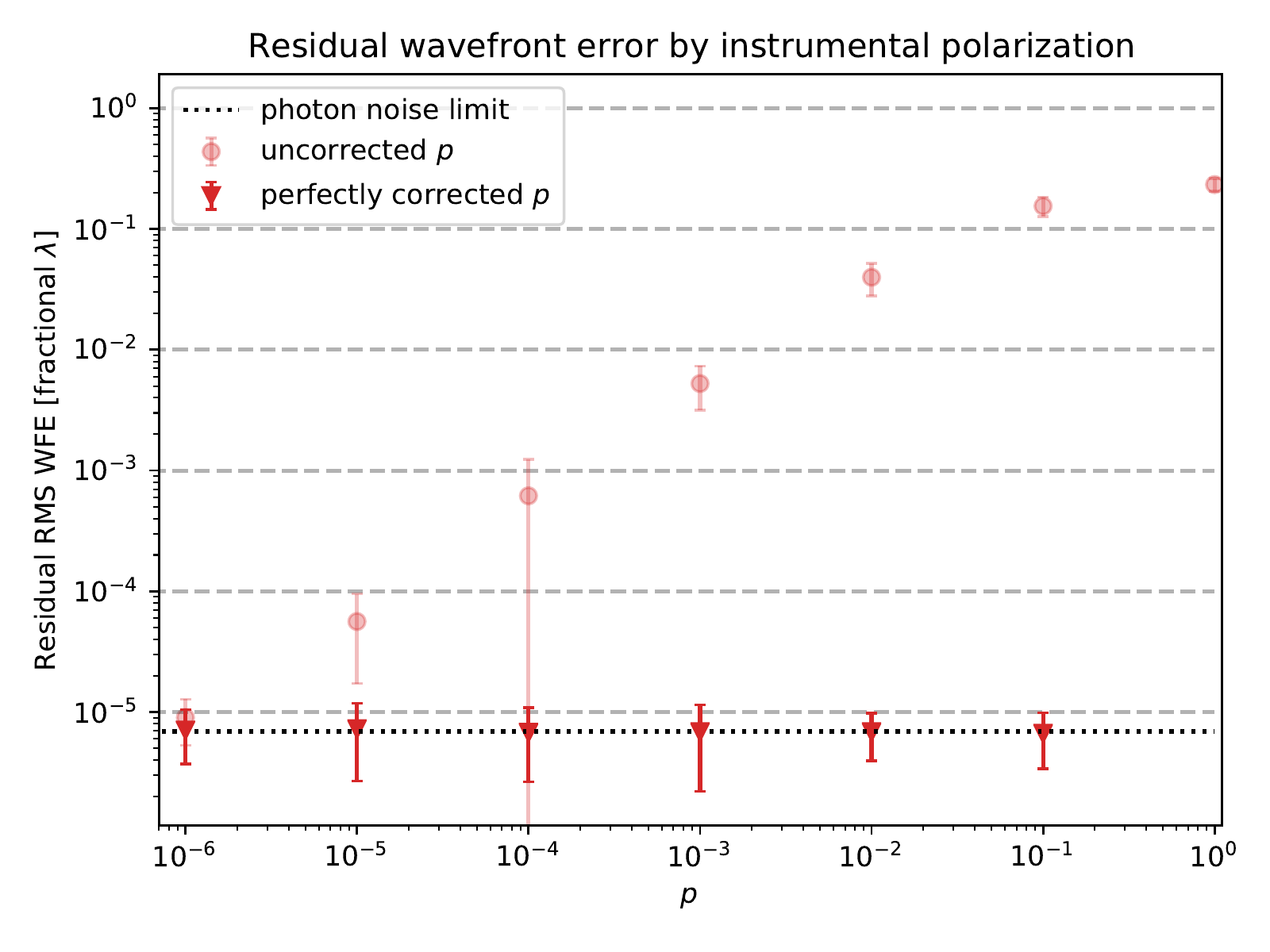}
\caption{
Performance of the wavefront sensing with instrumental polarization effects. 
The error bars show the $1\sigma$ deviation over the 100 random wavefront aberration instances. 
The circles show that data points where the strength of instrumental polarization ($p$) was not corrected and the triangles show the data points where $p$ was corrected. 
The dotted lines shows the photon noise limit, which was introduced to prevent the residual wavefront error reaching numerical noise. 
}
\label{fig:degree_of_polarization}
\end{figure}
As discussed and analytically studied in \autoref{subsec:instrumental_polarization}, uncorrected instrumental polarization can impact the wavefront estimation performance of the PESCC as the main peak in the OTFs does not, thus, completely cancel in the $\Delta\text{OTF}$.
Here, we quantify the effects of uncorrected and perfectly corrected instrumental polarization on the wavefront sensing. 
In \autoref{fig:degree_of_polarization}, we plot the effects of increasing $p$ versus the wavefront sensing performance. 
We show the case of corrected and uncorrected $p$ and we added photon noise ($10^{16}$ photons before the coronagraph) because, otherwise the corrected dots would be at the numerical noise limit.   
This shows that when $p$ is accurately known and corrected for, the detrimental effects can be completely amended.  
However, when $p$ is left (even partially) uncorrected, it significantly impacts the wavefront sensing. 
For $p\approx2\cdot10^{-3}$, the residual RMS WFE is $\sim10^{-2}$ $\lambda$. 
This shows that the application of the $p$ correction, suggested in \autoref{subsec:instrumental_polarization}, is important.  
{For SPHERE/IRDIS, the uncorrected level instrumental polarization is at the order of $\sim$$10^{-2}$ \citep{van2020polarimetric}.}
\subsubsection{Polarization leakage}\label{subsubsec:imperfect_polarizer_wfs}
\begin{figure}
\centering
\includegraphics[width=\hsize]{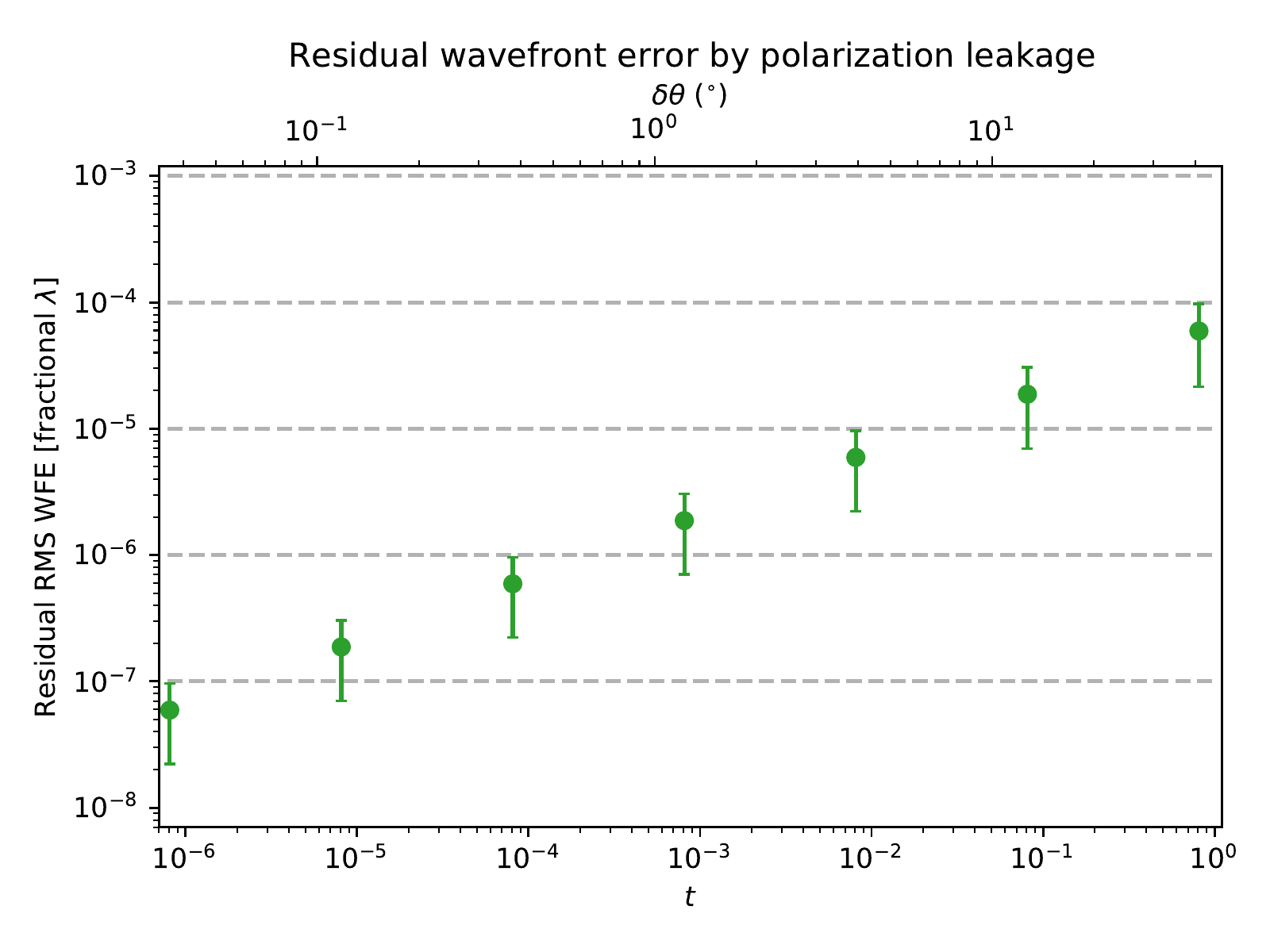}
\caption{
Performance of the wavefront sensing with polarization leakage.
The error bars show the $1\sigma$ deviation over the 100 random wavefront aberration instances. 
The bottom x-axis shows polarization leakage ($t$), and the upper x-axis shows the equivalent rotation offset ($\delta \theta$) between the RH polarizer and PBS. 
We note that the right most data point is sampled at $t=0.81$.
}
\label{fig:imperfect_polarizer}
\end{figure}
The polarizers are vital parts of the PESCC, and any imperfections will leak unwanted light into channel 2.
As discussed in \autoref{subsec:polarizer_leakage}, this will affect the wavefront estimation. 
This polarization leakage could be due to rotation offsets of this polarizer with regard to the polarizing beamsplitter or an imperfect blockage of the unwanted polarization state. 
Here, we simulate the polarization leakage and quantify its affect on the wavefront sensing performance of the PESCC. 
In \autoref{fig:imperfect_polarizer}, we plot the wavefront sensing performance as function of the polarizer leakage. 
It clearly shows that the polarizer leakage has a relatively minor impact on the wavefront sensing since, even for $t=0.81,$ the residual RMS WFE is still below $10^{-4}$ fractional $\lambda$.
 \subsubsection{Spectral resolution}\label{subsec:wfs_broadband}
 \begin{figure}
\centering
\includegraphics[width=\hsize]{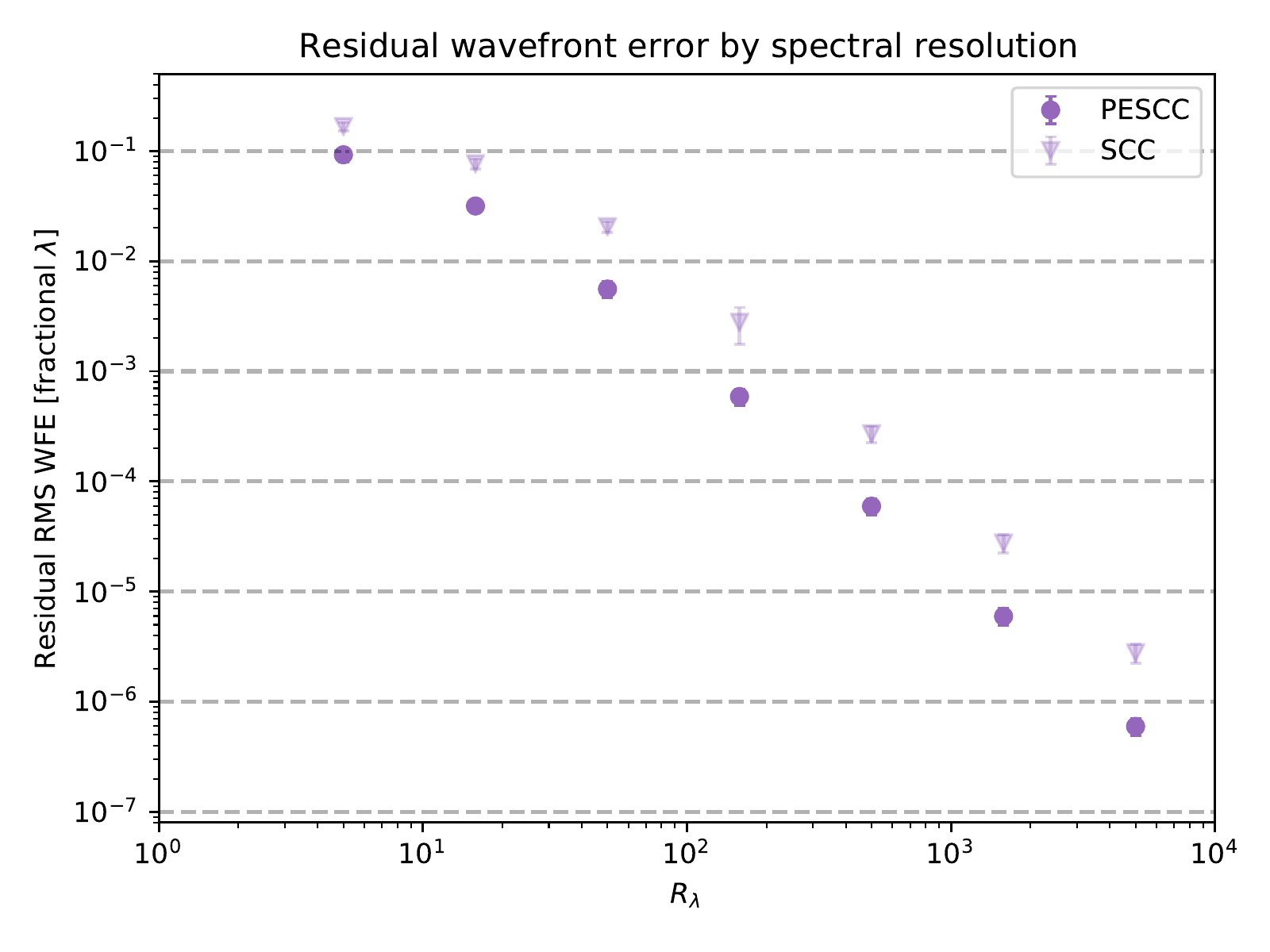}
\caption{
Performance of the wavefront sensing with varying spectral resolution ($R_{\lambda}$).
The error bars show the $1\sigma$ deviation over the 100 random wavefront aberration instances. 
The circles show the performance of the PESCC and the triangles the performance of the SCC. 
}
\label{fig:broadband_effects}
\end{figure}
Astronomical observations always have a finite $R_{\lambda}$, and, as was analytically studied in \autoref{subsec:bandwidth_limitations}, this affects the performance of the PESCC. 
Specifically, it was determined when \autoref{eq:spectral_resolution_SCC} and \autoref{eq:spectral_resolution_PESCC} are not satisfied, respectively, the SCC and PESCC, accurate wavefront sensing in the entire control region of the DM is not possible. 
Here we simulate the effects of spectral resolution on the wavefront sensing. 
The broadband effects are simulated by sampling seven wavelengths over the wavelength range defined by the spectral resolution, calculating the PSF for each wavelength, and incoherently adding the resulting PSFs.  
Due to the spectral effects, the sidebands in the $\Delta \text{OTF}$ are smeared, affecting the wavefront information.
We use the position and size of the RH at the central wavelength to generate an aperture that is applied to the $\Delta \text{OTF}$ for wavefront sensing. 
In \autoref{fig:broadband_effects}, the results of the simulation are shown. 
It shows that $<10^{-2}$ $\lambda$ RMS WFE is achieved for $R_{\lambda} \approx 30$ for the PESCC and $R_{\lambda} \approx 90$ for the SCC. 
\subsection{Wavefront sensing \& control}\label{subsec:wavefront_control_sims}
\begin{figure*}
\centering
\includegraphics[width=17cm]{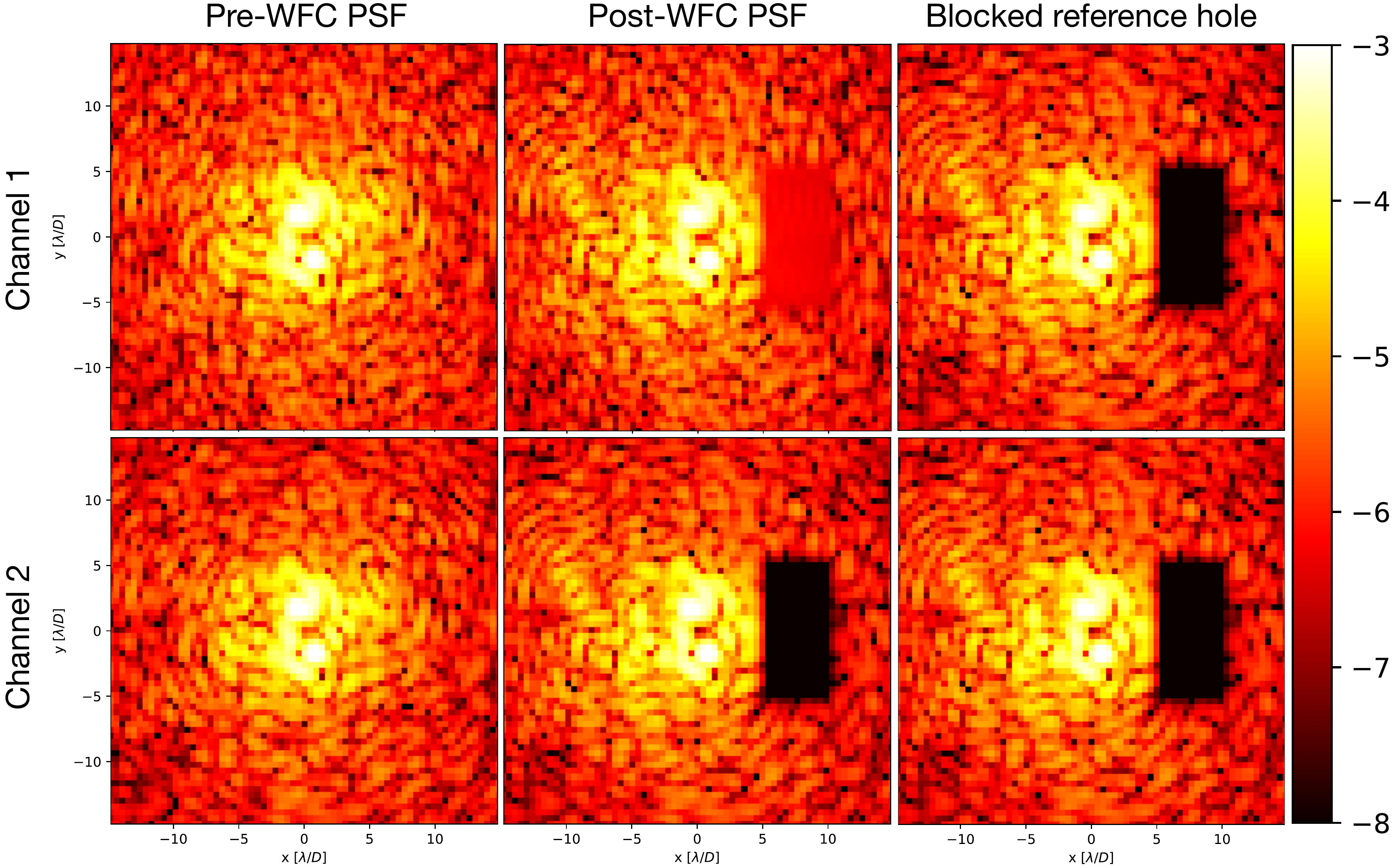}
\caption{WFC example with the PESCC without noise sources present.
             The subfigures in the two rows show the PSFs of the two channels.
             The columns show, respectively, the PSFs before WFC, the PSFs after twenty iterations of the WFC, and the PSFs after the WFC with the RH blocked.   
             The colorbar shows the intensity in logarithmic scale and is equal for all subfigures.  
             }
\label{fig:wavefront_control_example}
\end{figure*}

\begin{figure}
\centering
\includegraphics[width=\hsize]{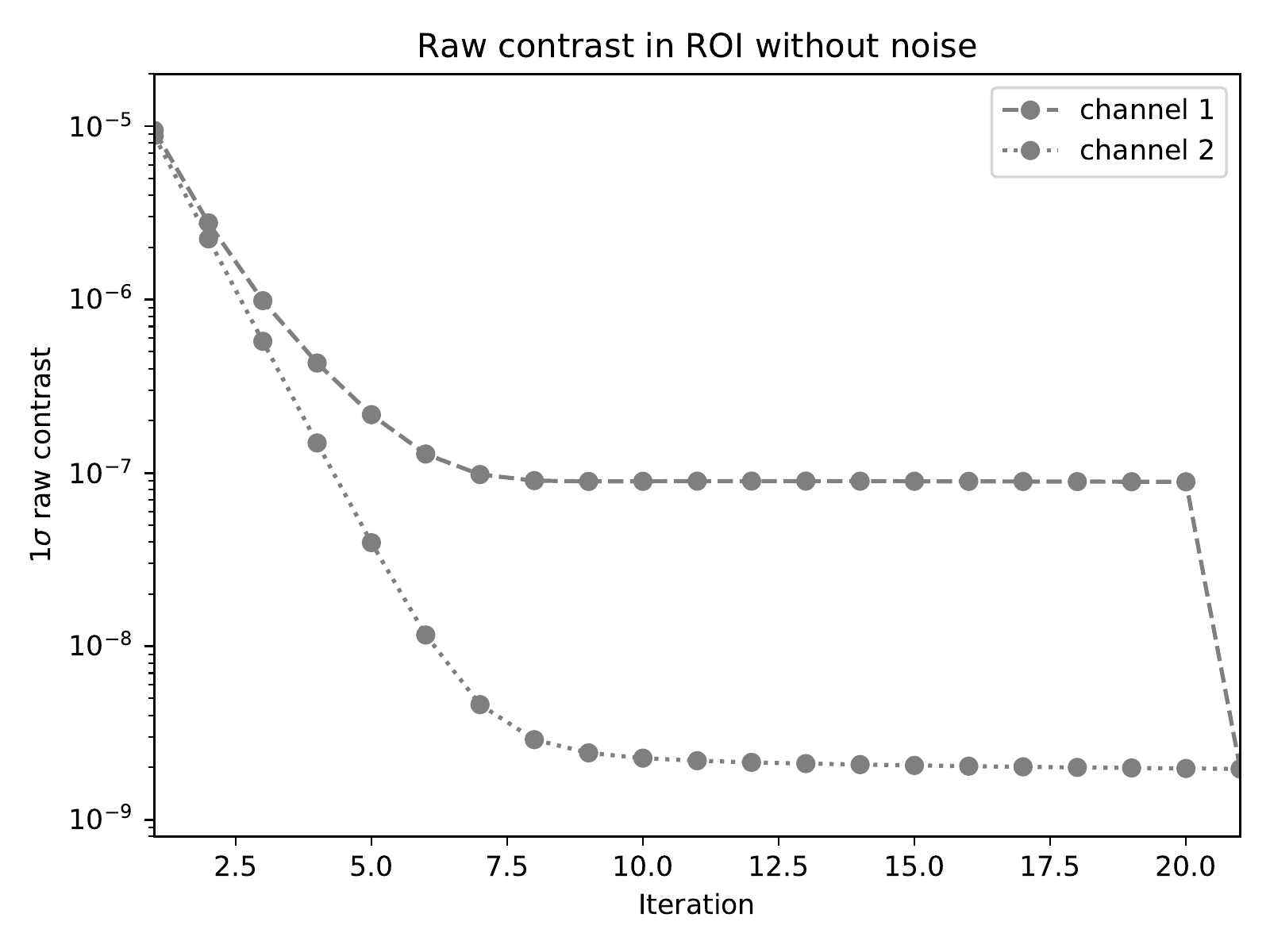}
\caption{Raw contrast as function of iterations in the ROI for WFC example with noise sources. 
             At iteration 21, the RH is blocked to show that the dark holes reach similar raw contrasts. 
             }
\label{fig:raw_contrast_perfect}
\end{figure}
In this subsection, we study how the noise sources in \autoref{subsec:wavefront_sensing_sims} affect the wavefront sensing and control when the goal is to cancel residual starlight in the region of interest (ROI). 
When the goal is to minimize the electric field of the starlight in the ROI, it was shown \citep{mazoyer2014high} that minimizing $I_{sb}$ (\autoref{eq:I_sb_estimation}) is equivalent. 
We assume that we are in the small phase regime and can therefore say that there is a linear relationship between DM actuation and changes in $I_{sb}$. 
The response matrix, which dictates this linear relationship, can now be calibrated by actuating modes on the DM (actuator pokes or sine/cosine modes) and recording the changes in $I_{sb}$. 
We find the focal-plane response to DM mode $i$ by:
\begin{equation}
\Delta I^i_{sb} = \frac{I^i_{sb+} - I^i_{sb-}}{2 a^2}, 
\end{equation}  
with $a$ the poke amplitude that was set to $10^{-3} \cdot \lambda$, and $I^i_{sb+}$ and $I^i_{sb-}$ the images flattened to 1D vectors that corresponds to the positive and negative actuations of the mode, respectively. 
The response matrix $R$ is then constructed by stacking the responses to the $N$ modes that are controlled: 
\begin{equation}
R = 
\begin{pmatrix}
\mathbb{R} \{ \Delta I^1_{sb}  \} & \mathbb{I} \{ \Delta I^1_{sb}  \}  \\ 
\vdots & \vdots\\ 
\mathbb{R} \{ \Delta I^N_{sb}  \} & \mathbb{I} \{ \Delta I^N_{sb}  \}  \\
\end{pmatrix}^\mathsf{T},
\end{equation}
with $\mathbb{R} \{ \cdot  \}$ and $\mathbb{I} \{ \cdot  \}$ the real and imaginary components. 
In the simulations there is only one DM, which is set in a pupil-plane and, therefore, we can only hope to correct for phase and amplitude errors in a one-sided dark hole. 
Therefore, we chose a ROI given by  $5 \ \lambda/D < x < 10 \ \lambda / D$ and  $-5 \ \lambda/D < y < -5 \ \lambda / D$. 
We use a sine/cosine mode basis to directly probe this region \citep{poyneer2005optimal} to calibrate the response matrix. 
The control matrix $C$ is then calculated by inverting the response matrix using the singular-value decomposition method with Tikhonov regularization. 
In closed-loop operation for the wavefront control (WFC), we use a simple integral controller with a loop gain of 0.5. \\
\indent As a comparison to WFC tests with the various noise sources, we first simulate the WFC for 20 iterations without any noise sources present. 
The PSFs of this test are shown in \autoref{fig:wavefront_control_example}. 
It shows that a dark hole is generated after the WFC in the ROI. 
In channel 1, the intensity of the RH is clearly visible as it limits the achieved contrast. 
When the RH is blocked, the contrast in the ROI is the same for both channels. 
The convergence of the algorithm is shown in \autoref{fig:raw_contrast_perfect}. 
The contrast of the two channels is plotted, which shows that the channel with the RH PSF plateaus at $\sim10^{-7}$ before the RH is blocked, and the other channel converges to $\sim2 \cdot 10^{-9}$ within ten iterations.
When the RH is blocked at iteration 21, the contrast in channel 1 also converges to $\sim2 \cdot 10^{-9}$. \\   
\indent The results presented in \autoref{fig:raw_contrast_perfect} serve as benchmark for the test with noise sources to quantify the performance loss. 
For all the noise sources tested in \autoref{subsec:wavefront_control_sims}, we use the values that give approximately a $10^{-1}$, $10^{-2}$, $10^{-3}$ fractional $\lambda$ RMS WFE to test with WFC. 
 \subsubsection{Photon noise performance}
 \begin{figure}
\centering
\includegraphics[width=\hsize]{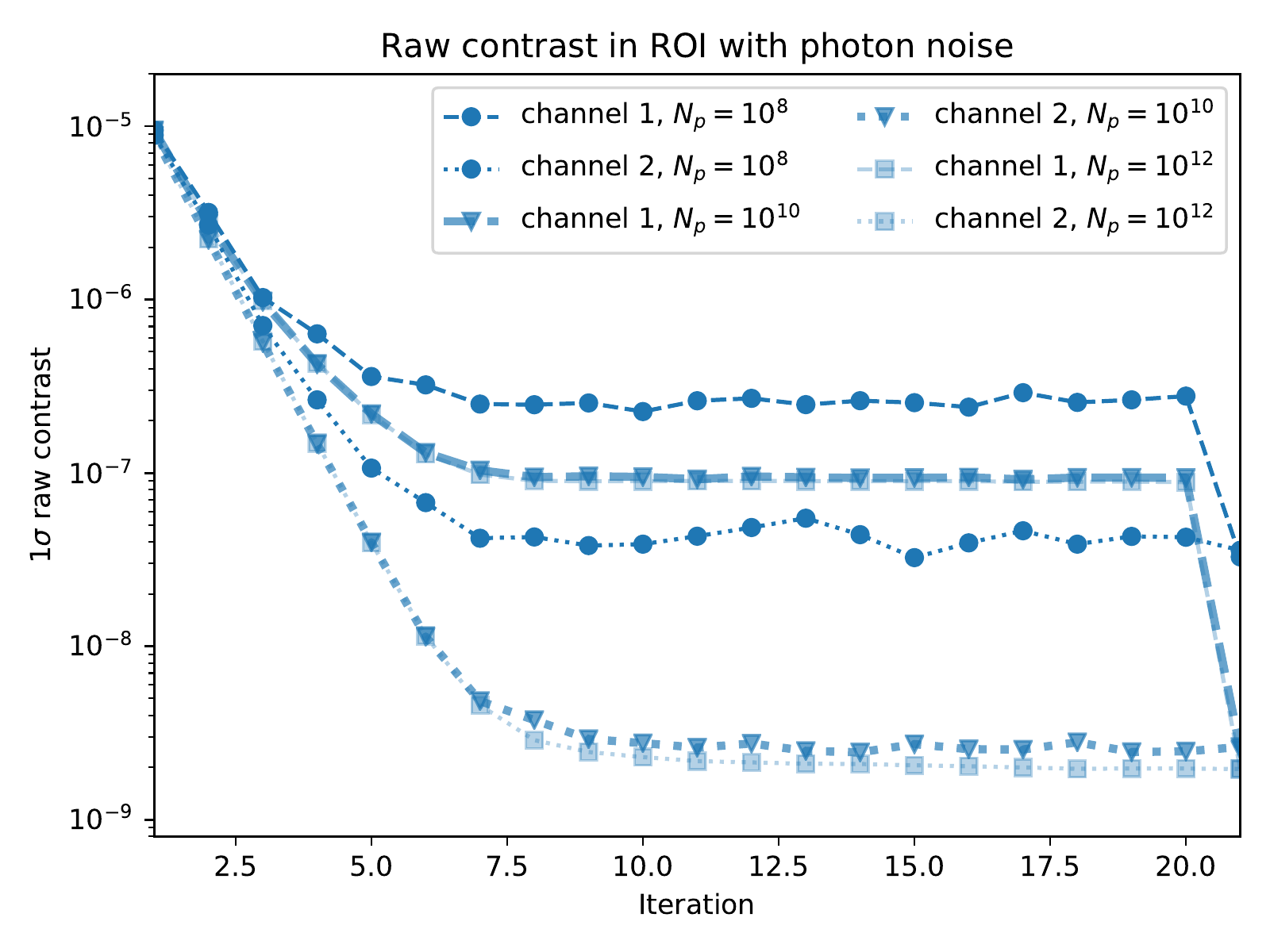}
\caption{Raw contrast as function of iterations for the system with photon noise.
             The response matrix is acquired without photon noise. 
             At iteration 21, the RH is blocked to compare the raw contrasts in the two channels.
             }
\label{fig:raw_contrast_photon_noise}
\end{figure}
Here, we test how the wavefront control converges under photon noise. 
We assume that the response matrix is calibrated by observing a very bright source (e.g., an internal source within the instrument) such that photon noise is irrelevant, that is, we did not simulate photon noise while acquiring the response matrix. 
In \autoref{fig:raw_contrast_photon_noise}, the results are presented.
For $10^8$ photons per exposure the wavefront control converges to $\sim4 \cdot 10^{-8}$ contrast. 
With $10^{10}$ photons, the contrast is close to that of the perfect system, at $\sim3 \cdot 10^{-9}$. 
Then for $10^{12}$ photons, the contrast that is reached is that of the perfect system, $\sim2 \cdot 10^{-9}$.  
When considering the current internal near-infrared (NIR) camera and $\Delta \lambda = 50$ nm filter at 1550 nm in SCExAO (\citealt{jovanovic2015subaru}; \citealt{lozi2018scexao}), which is located at the 8 meter Subaru telescope on Maunakea, these photon numbers correspond to $\sim 2$ Hz WFC loop speed on a $m_H = 6$, $m_H = 1$, and $m_H = -4$ target, respectively.  
 \subsubsection{Differential aberrations}
\begin{figure}
\centering
\includegraphics[width=\hsize]{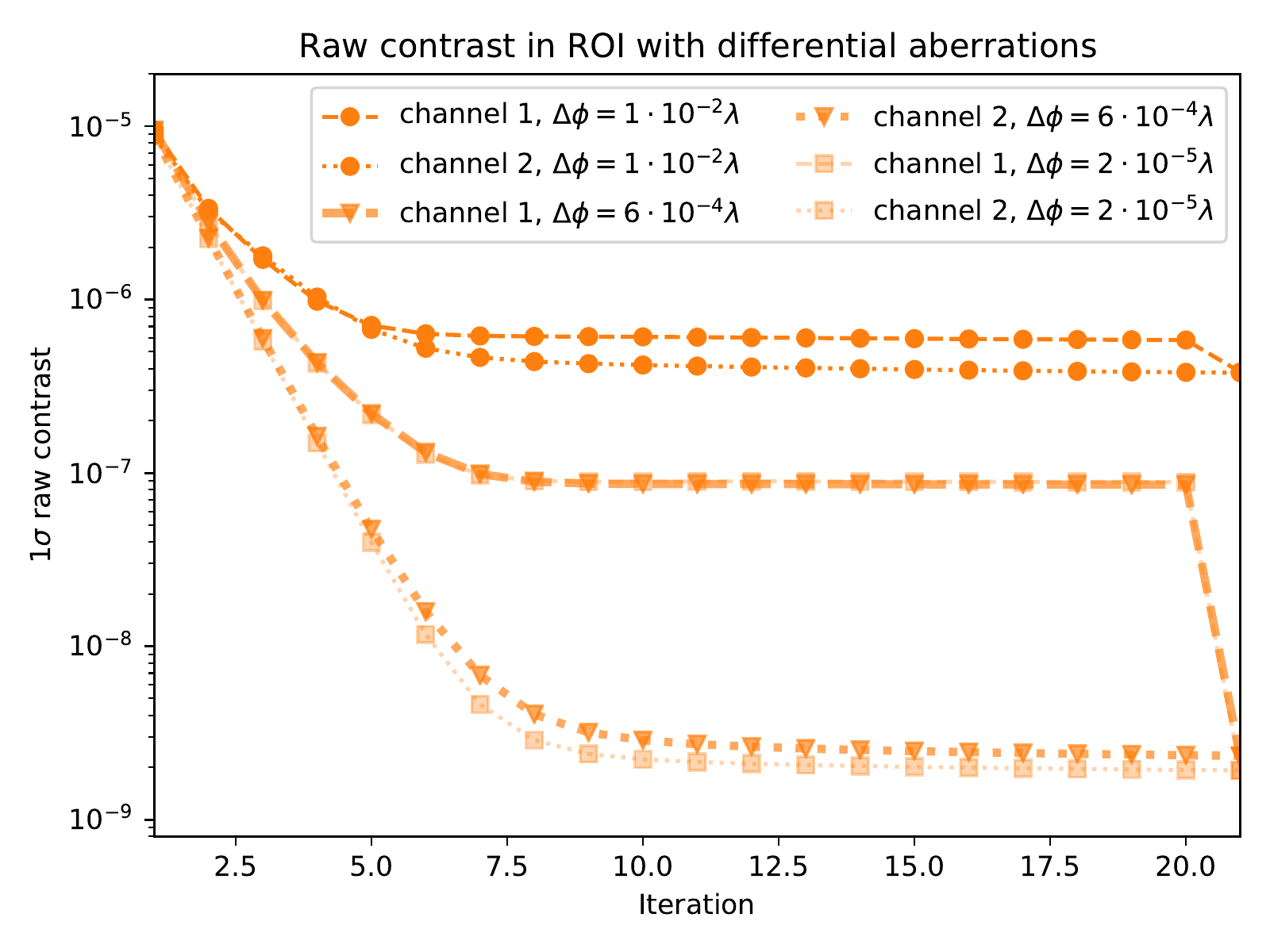}
\caption{Raw contrast as function of iterations for the system with different levels of RMS differential aberrations ($\Delta \phi$) between the beams.
             The response matrix is acquired with the differential aberrations present. 
             At iteration the RH is blocked to compare the raw contrast in the two channels.  
             }
\label{fig:raw_contrast_differential_aberrations}
\end{figure}
As shown in \autoref{subsubsec:differential_aberrations_wfs}, differential aberrations between the two beams after the PBS severely affect the wavefront sensing. 
Here, we quantify to what level it limits the WFC.  
When calibrating the response matrix, the differential aberrations are included as they are expected to be always present in the system. 
In \autoref{fig:raw_contrast_differential_aberrations}, the convergence of the WFC under various levels of differential aberration is shown. 
For $\Delta \phi = 10^{-2}$ $\lambda,$ the PESCC converges to a contrast of $\sim4 \cdot 10^{7}$.
With $\Delta \phi = 6 \cdot 10^{-4}$ $\lambda$ it converges close to the benchmark performance at $\sim 3 \cdot 10^{-9}$ contrast.
For $\Delta \phi = 2 \cdot 10^{-5}$ $\lambda$ the system converges to the benchmark system results at $\sim 2 \cdot 10^{-9}$. 
To put these values into perspective, the SCExAO/CHARIS polarization mode \citep{lozi2019new} has $\sim7 \cdot 10^{-3}$ waves of differential aberrations\footnote{As derived from a Zemax file provided by T. Groff.}, and SPHERE/IRDIS was built with $\sim6 \cdot 10^{-3}$ waves of differential aberrations \citep{dohlen2008infra} (both values calculated at $\lambda = 1600$ nm). 
This means that if PESCC were implemented at either of these systems, it would converge to a 1$\sigma$ raw contrast between $\sim4 \cdot 10^{7}$ and $\sim 3 \cdot 10^{-9}$, probably closer to the former. 
\subsubsection{Instrumental polarization} 
\begin{figure*}
\centering
\includegraphics[width=17cm]{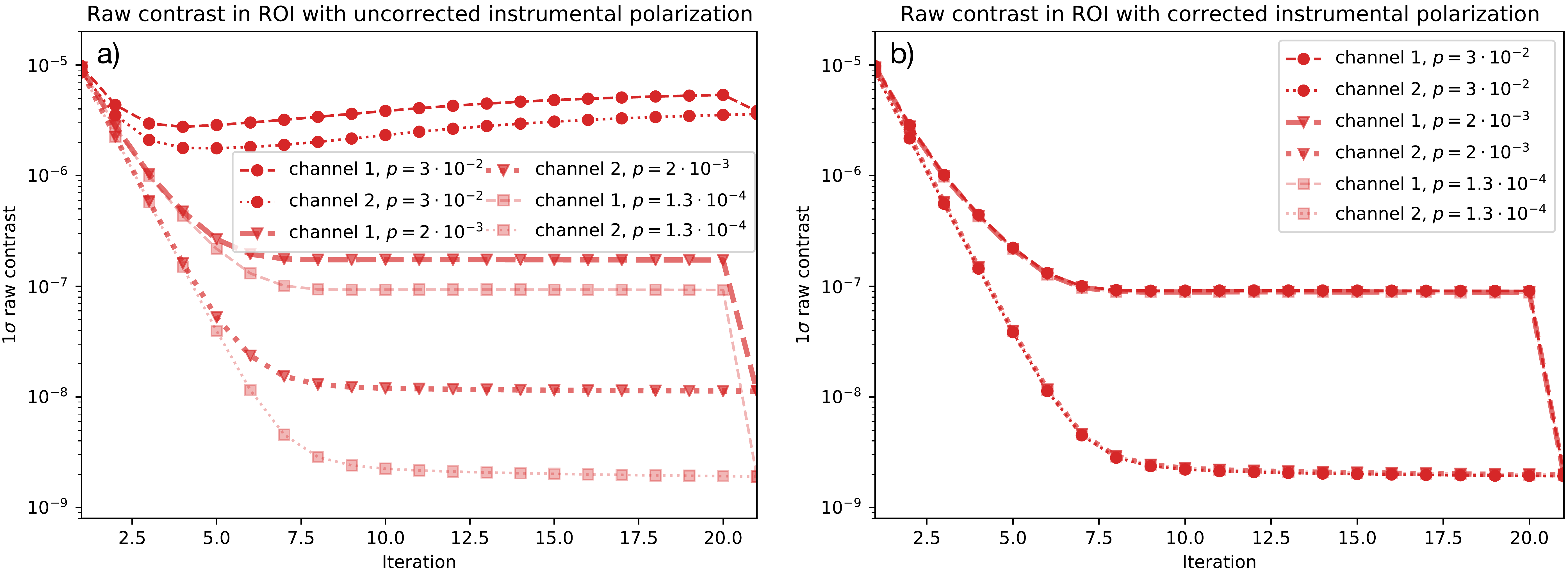}
\caption{Raw contrast as function of iterations for the perfect system with instrumental polarization ($p$) effects.
             At {the last} iteration, the RH is blocked to compare the raw contrast in the two channels.
             (a) During the acquisition of the response matrix and the closed loop tests the effects of $p$ were not corrected. 
             (b) During all steps of the response matrix acquisition and WFC tests, $p$ was corrected.
             All three tested levels of $p$ now overlap and {reach} the performance {level} of the benchmark tests.  
             }
\label{fig:raw_contrast_instrumental_polarization}
\end{figure*}
Here, we test the effect of (un)corrected instrumental polarization on the WFC with the PESCC. 
When calibrating the response matrix the instrumental polarization effects are included when there is no correction of $p$ in post-processing.
When the instrumental polarization is corrected, the correction is also included when calibrating the response matrix.  
In \autoref{fig:raw_contrast_instrumental_polarization} a we show the WFC results when the instrumental polarization is not corrected. 
For $p=3\cdot10^{-2}$, the loop is not stable because the contrast first increases, and then decreases. 
The final contrast achieved is $\sim4 \cdot 10^{-6}$, which is only a slight improvement from the initial contrast. 
When $p=2\cdot10^{-3}$, the system converges to $\sim10^{-8}$ contrast.
Then, for $p=1.3\cdot10^{-4}$ the contrast achieved is $\sim2 \cdot 10^{-9}$, equal to the benchmark results. 

In \autoref{fig:raw_contrast_instrumental_polarization} b the polarization effects are corrected during WFC by the method presented in \autoref{subsec:instrumental_polarization}.
It shows that for all cases, the WFC converges to the contrast of the system without noise. 
Uncorrected instrumental polarization at SPHERE/IRDIS is at a level of $p\approx 10^{-2}$, and when corrected, using a detailed instrument polarization model, reaches $p \leq 10^{-3}$ \citep{van2020polarimetric}.
As shown in the results of \autoref{fig:raw_contrast_instrumental_polarization} a, this means that the instrumental polarization have to be corrected as, otherwise, the loop would be unstable and diverge. 
When the instrument polarization model is used and $p$ is corrected to a level of $\sim10^{-3}$, then the WFC will converge to $\sim10^{-8}$ contrast.
\subsubsection{Polarization leakage}
\begin{figure}
\centering
\includegraphics[width=\hsize]{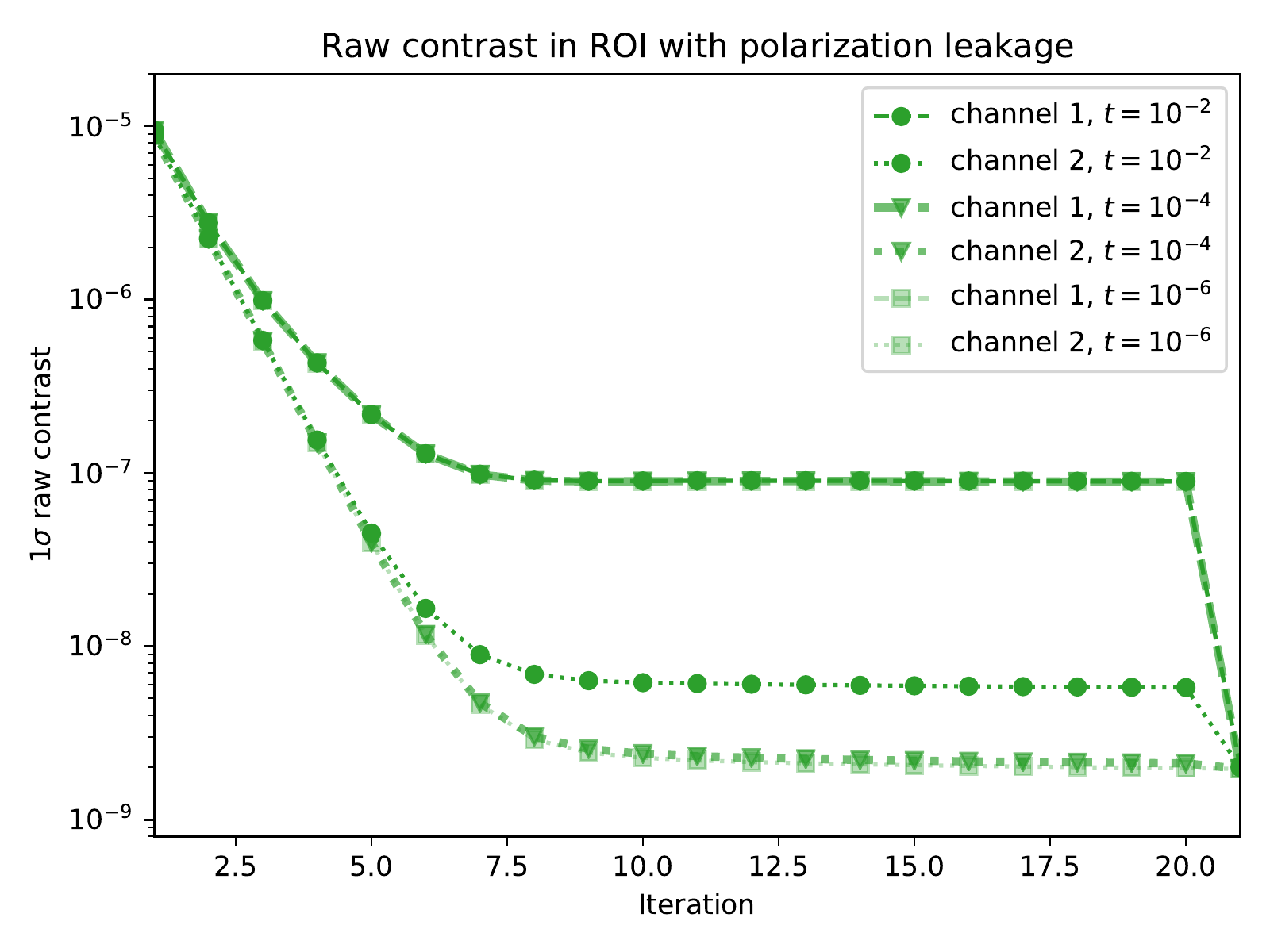}
\caption{Raw contrast as function of iterations for the system with polarizer leakage.
             The response matrix is acquired with the polarizer leakage.
             At iteration 21, RH is blocked to compare the raw contrast between the two channels.  
             }
\label{fig:raw_contrast_polarizer_leakage}
\end{figure}
We investigate the effects of polarizer leakage on the WFC. 
As with the previous subsections, we include the polarizer leakage in the response matrix calibration.  
The results presented in \autoref{subsubsec:imperfect_polarizer_wfs} show that the wavefront error is never affected more than on the level of a $10^{-4}$ fractional $\lambda$ RMS WFE. 
Therefore, we decided to test $t = 10^{-2}, 10^{-4}, 10^{-6}$.
The results are shown in \autoref{fig:raw_contrast_polarizer_leakage}. 
It shows that for $t=10^{-3}$, the contrast in channel 2 initially does not converge to the benchmark contrast. 
When the RH is blocked, then both channels converge to $\sim2 \cdot 10^{-9}$, which shows that channel 2 was limited by leakage from the RH PSF. 
This proves that the WFC itself is not limited by polarizer leakage, but that the contrast in the DH could be limited by leakage from the polarizer. 
As discussed in \autoref{subsec:polarizer_leakage}, polarization leakage is expected to be $t=10^{-3}-10^{-4}$, and $t\leq10^{-5}$ for the RH polarizer and PBS, respectively, and are not expected to to have an impact on WFC. 
When the RH polarizer and PBS are misaligned by $5^{\circ}$ the leakage is $\sim2\cdot10^{-2}$, which is also on a level that does not impact the WFC, but the contrast in channel 2 will then be limited by RH PSF to $\sim7 \cdot 10^{-9}$. 
 \subsubsection{Spectral resolution}
\begin{figure}
\centering
\includegraphics[width=\hsize]{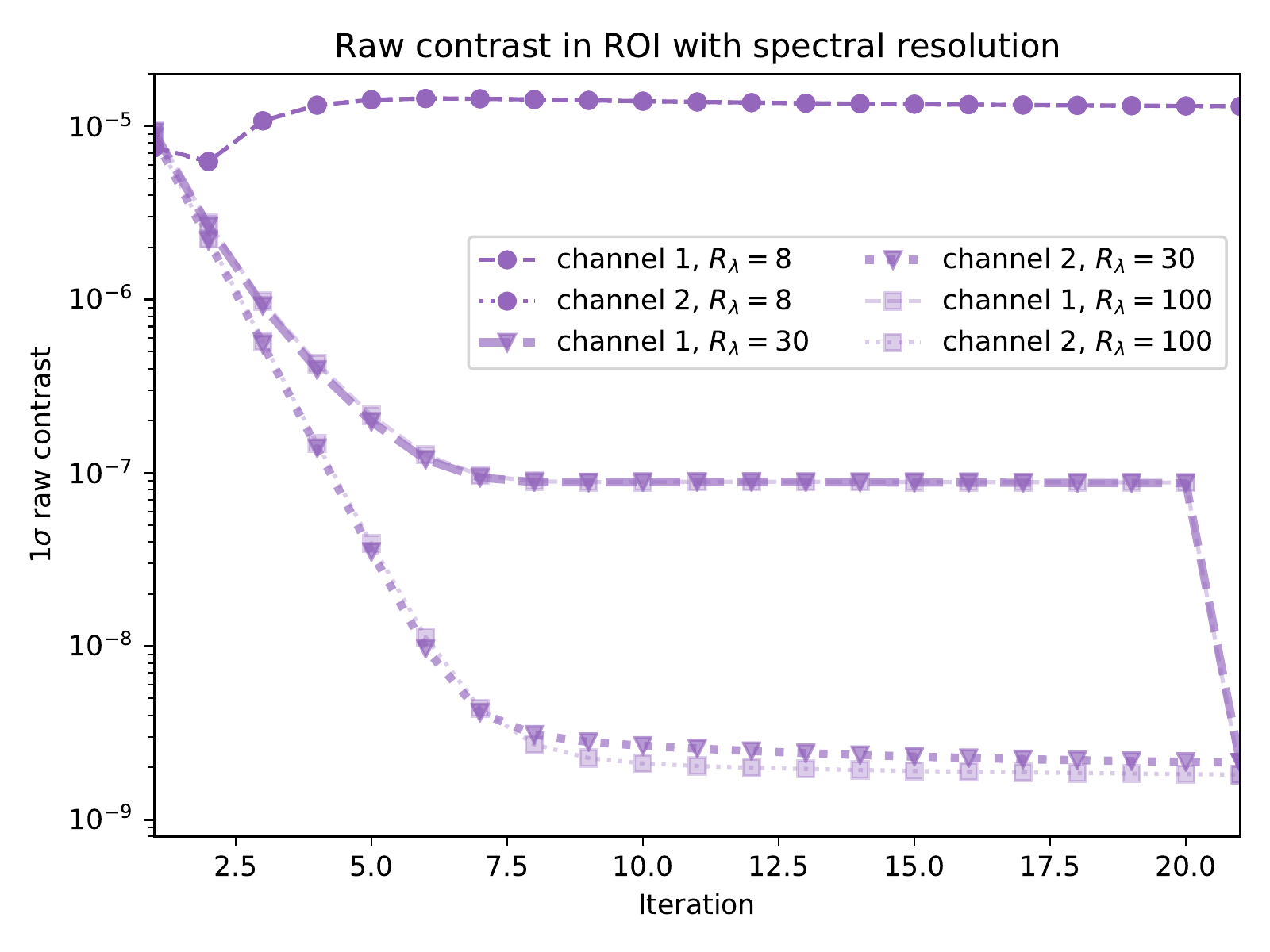}
\caption{Raw contrast as function of iterations for the system with broadband effects.
             The response matrix is acquired with the broadband effects included.
             At iteration 21, the RH is blocked to compare the raw contrast in the two channels.
             }
\label{fig:raw_contrast_broadband_effects}
\end{figure}
Here, we simulate the effects of spectral resolution on WFC.
The broadband effects are included when the response matrix is measured and the wavefront measurements are identical to \autoref{subsec:wfs_broadband}. 
Similarly as before, we sample seven wavelengths of the spectral band and add the resulting PSFs to get the broadband PSF. 
In \autoref{fig:raw_contrast_broadband_effects}, we show the WFC results for various $R_{\lambda}$. 
It shows that for $R_{\lambda} = 8,$ the WFC control diverges and that the contrast in the ROI becomes worse.
For $R_{\lambda} = 30$ and $R_{\lambda} = 100,$ the WFC converges (close) to the contrast achieved by the system without noise sources. 
The tested spectral resolutions are equivalent to filters with bandwidths of $0.13 \cdot \lambda_0$, $0.03 \cdot \lambda_0$, and $0.01 \cdot \lambda_0$.  
Therefore, as shown in \autoref{fig:raw_contrast_broadband_effects}, the PSECC does not work with the broadband photometric filters, but with narrowband filters that have $\delta \lambda \leq 0.03 \cdot \lambda_0$ it will be able to run a WFC effectively. 
Operation of the PESCC with an integral field spectrograph (IFS) would be an ideal solution as it provides relatively narrowband images over broad wavelength ranges. 
SCExAO/CHARIS \citep{groff2017first} offers a low-resolution mode at $R_{\lambda} = 18$ and high-resolution modes at $R_{\lambda}\approx70$.
The PESCC is able to operate with the high-resolution modes. 
And SPHERE/IFS operates either with $R_{\lambda} = 30$ \citep{claudi2008sphere} or $R_{\lambda} = 50$ \citep{mesa2015performance}, which means that the PESCC can operate with both modes. 
\subsubsection{Combined effects}
\begin{figure}
\centering
\includegraphics[width=\hsize]{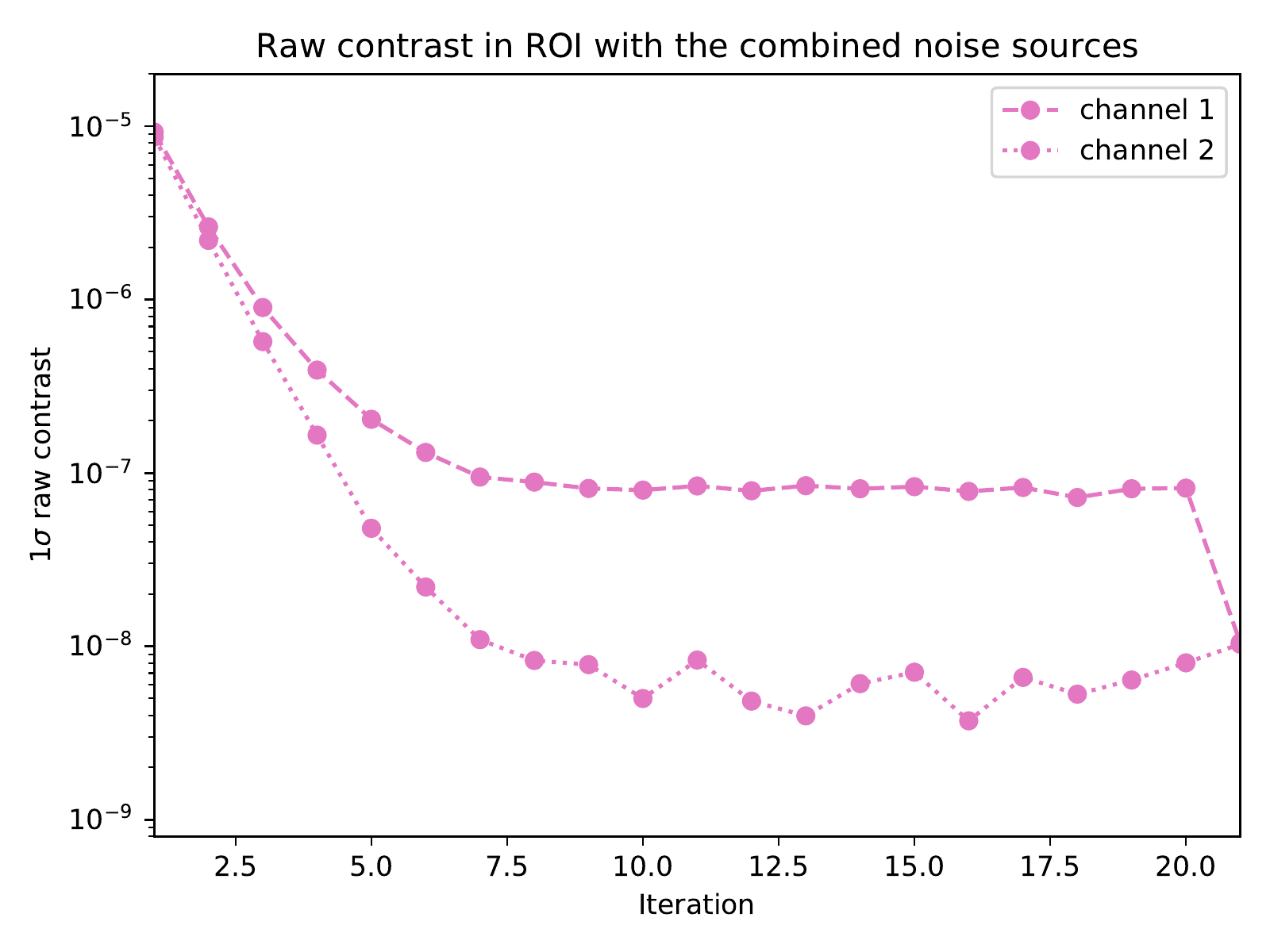}
\caption{Raw contrast as function of iterations for the system with all effects.
             The response matrix is acquired with all effects included.
             At iteration 21, the RH is blocked to compare the raw contrasts in the two channels. 
             }
\label{fig:raw_contrast_combined_effects}
\end{figure}
Here, we combine the tested noise sources in one simulation to investigate whether these noise sources interact with each other. 
For every noise source, we select the level that gives a $10^{-2}$ $\lambda$ WFE, as described in \autoref{subsec:wavefront_sensing_sims}.  
All the noise sources, except the photon noise, are included when calibrating the response matrix. 
During the wavefront sensing step, the instrumental polarization is measured and corrected in the frames. 
The results are presented in \autoref{fig:raw_contrast_combined_effects}. 
This shows that the WFC control converges to a contrast of $\sim10^{-8}$, which is very similar to the raw contrast achieved when there were uncorrected $p$ effects at $p \sim 2 \cdot 10^{-3}$, roughly an order of magnitude worse than the performance under the other individual noise sources. 
This is caused by the differential aberrations, which introduce intensity differences between the OTFs of the two channels at locations where $p$ is measured. 
These intensity differences are not caused by $p$ effects and, therefore, lead to incorrect $p$ estimates. 
\subsection{Coherence differential imaging}\label{subsec:CDI_sims}
\begin{figure*}
\centering
\includegraphics[width=17cm]{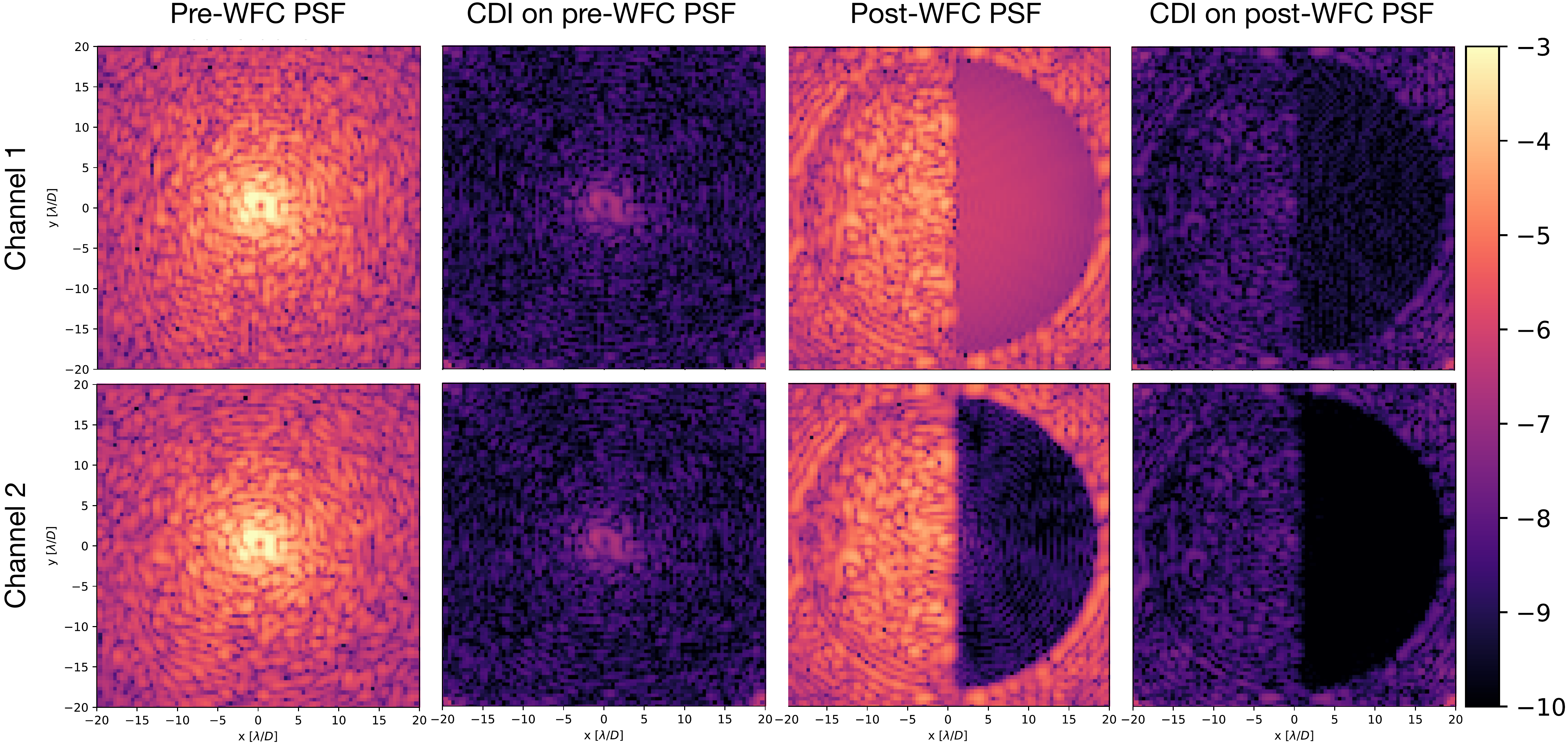}
\caption{WFC and CDI  example with the PESCC for a system without noise source other than wavefront aberrations.
             The two rows show subfigures of the two channels. 
             The columns show the PSFs before WFC, the PSFs before WFC and after CDI, the PSFs after WFC, and the PSFs after WFC and CDI. 
             The colorbar shows the intensity in logarithmic scale and is equal for all subfigures.  
             }
\label{fig:CDI_example}
\end{figure*}
\begin{figure}
\centering
\includegraphics[width=\hsize]{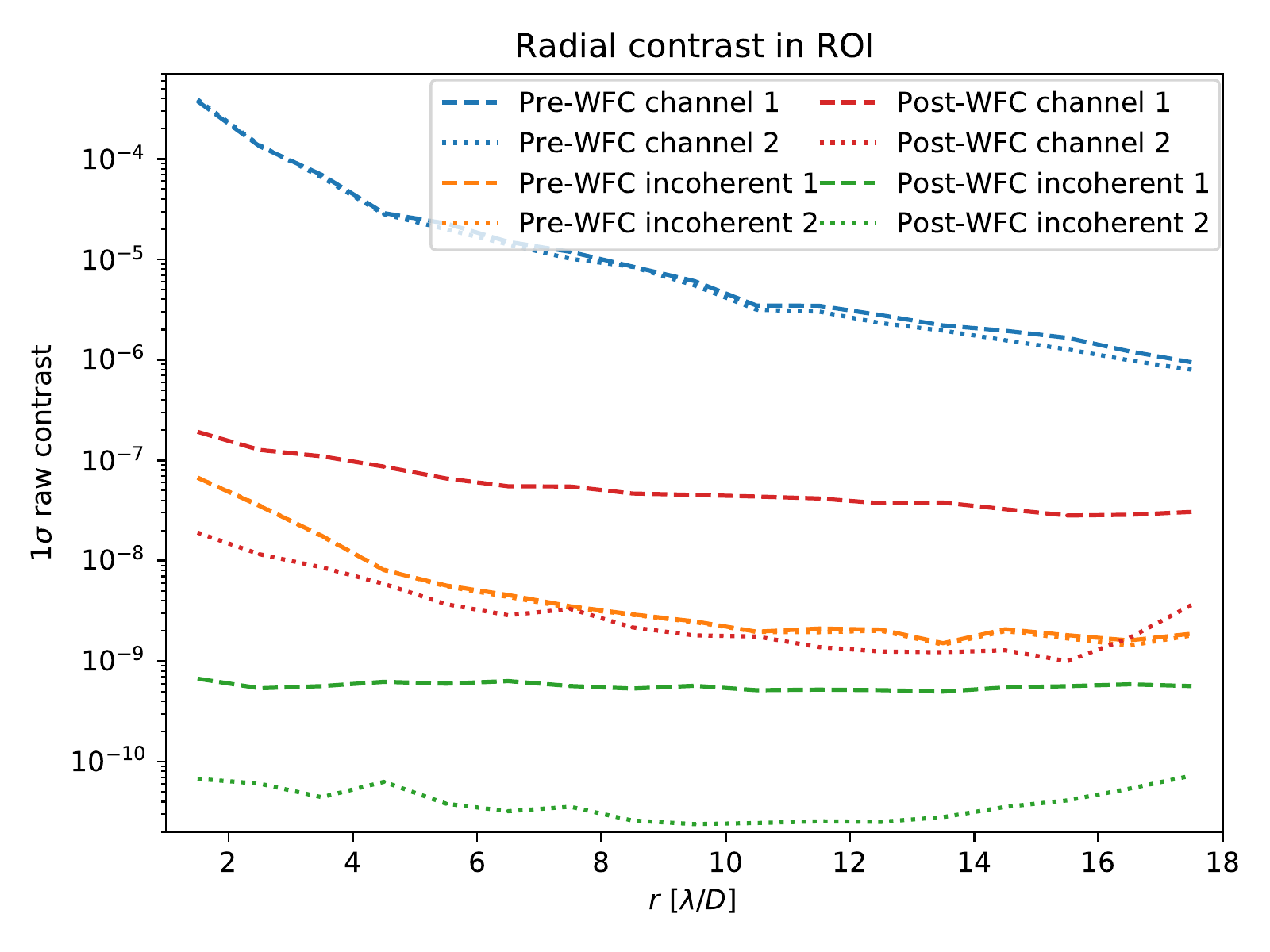}
\caption{Radial contrast in the ROI for the CDI simulation.
Shown are the $1\sigma$ raw contrast curves before and after WFC and CDI.  
             }
\label{fig:radial_contrast}
\end{figure}
In this subsection, we study the improvements in contrast that CDI could bring with the PESCC, which was developed in \autoref{subsec:CDI}. 
As proof of principle, we simulate a monochromatic, idealized system without any noise sources other than wavefront aberrations. 
The parameters as presented in \autoref{tab:simulation_parameters} are used for the simulation. 
We compare the CDI before and after the WFC.  
We aim to minimize the starlight in the ROI, defined by $1 \ \lambda/D > r > 18 \ \lambda / D$ and $x>0$.
{The ROI is larger than what was used in \autoref{subsec:wavefront_control_sims} to show that the PESCC is not limited to a dark hole size.}
Unlike in \autoref{subsec:wavefront_control_sims}, we do not block the RH after the final WFC step.
This is to estimate the effect of subtracting the terms involving the RH PSF. 
In \autoref{fig:CDI_example}, the PSFs are shown before and after the WFC as well as after the CDI. 
It shows that the WFC improves the contrast in the ROI, as it is intended to do, while the CDI improves the entire FOV. 
Furthermore, it shows that CDI is not able to completely remove the PSF and bring the contrast to numerical noise. 
This is likely due to numerical artifacts. 
The PSF of the RH is clearly visible in the post-WFC PSF of channel 1, and is largely removed after the final CDI step. 
A radial profile of the contrast in the ROI is shown in \autoref{fig:radial_contrast}. 
It shows that the initial contrast in the ROI is between $\sim 4 \cdot 10^{-4}$ and $\sim10^{-6}$. 
When performing CDI on the initial PSFs, the contrast is improved to $\sim6\cdot10^{-8}-3\cdot10^{-9}$, which is an increase of a factor of $\sim330-6600$. 
Following the WFC, the contrast becomes $\sim 2\cdot10^{-7} - 3\cdot10^{-8}$ for channel 1, which is limited by the RH PSF, and $\sim 2\cdot10^{-8} - 10^{-9}$ for channel 2. 
A subsequent CDI step brings this to $6\cdot10^{-10}$ for channel 1, and $\sim 3\cdot10^{-11} - 8\cdot10^{-11}$ for channel 2, which is a factor of $\sim100-300$ increase. 
\section{Discussion and conclusions}\label{sec:conclusions} 
In this work, we present the PESCC, a new variant of the SCC that features a linear polarizer in the RH.
When the two linear polarization states are subsequently separated by a polarizing beamsplitter, the focal-plane image of the polarization state let through by the RH polarizer {is} fringed, while the image of the orthogonal polarization state remains unmodulated by fringes. 
When the OTFs of these two images are subtracted, the sidebands containing wavefront information and the RH peak are revealed.  
These can be used for wavefront sensing and control in order to generate dark holes where the starlight is canceled by DM actuation. It can also be used for CDI, a post-processing technique that aims to remove all coherent light in the image. 
The PESCC has the great advantage that the RH can be placed right next to the pupil, which strongly reduces the constrains on the optics size as the total beam foot print is much smaller compared to the SCC.  
This makes it much easier to implement the PESCC in existing high-contrast imaging instruments such as VLT/SPHERE and Subaru/SCExAO. 
As focal-plane coronagraphs diffract more light close to the edge of the pupil, this has the added advantage that the PESCC has access to more light for wavefront sensing. \\

\indent {We simulated an idealized HCI system with static phase and amplitude wavefront aberrations generated by an out-of-plane phase aberration.
The system was operating at 1550 nm, consisting of a clear aperture, an idealized DM (e.g., no actuator cross-talk or quantization errors), with a 40$\times$40 square grid located in the pupil plane, a charge 2 vortex coronagraph, a (PE)SCC Lyot stop, and a polarizing beamsplitter and detector.} 
We found that the PESCC has $\sim$16 times more photons available than the SCC (this includes the 50\% throughput of the RH polarizer and polarizing beamsplitter). 
This was confirmed with additional simulations where we studied the sensitivity of the wavefront sensing with photon noise, as the PESCC reached a sensitivity $\sim4$ times higher than that of the SCC. 
This can either be used to increase the loop speed of the {WFC} or reach higher sensitivities in the wavefront sensing. 
The RH being closer the pupil also relaxes the focal-plane sampling and spectral resolutions constraints with regard to the SCC by a factor 2 and 3.5, respectively, to 2 pixels per $\lambda/D$ (which is a factor four gain in number of pixels) and $R_{\lambda}\approx24$ for an infinitely small RH.
The latter was confirmed with numerical simulations.
Another advantage is that the PESCC automatically estimates the RH {PSF}, enabling CDI post-processing for all science frames. 
Through idealized simulations, we have shown that CDI after WFC can reach a $1\sigma$ raw contrast of $\sim 3 \cdot 10^{-11} - 8 \cdot 10^{-11}$ between 1 and 18 $\lambda/D$. 
However, we found in the analytical and numerical studies that instrumental polarization and differential aberrations need to be tightly controlled for the PESCC to operate successfully. 
We have shown that it is possible to measure the degree of instrumental polarization in the OTFs of the two channels when the differential aberrations are not dominant, and this can subsequently be used to correct the images. 
If the differential aberrations dominate, then it is preferable to use an instrumental polarization model to predict $p$ \citep{van2020polarimetric}. 
Leakage from the RH polarizer was found not to affect the wavefront sensing significantly, but it does pollute the dark hole in the other channel. \\

\indent The simulations presented in \autoref{sec:simulations} were highly idealized and are, rather, more representative of space-based observatory conditions.
They mainly serve as a proof of principle and in a future work, we will use more realistic simulations to investigate the performance of the PESCC on ground-based systems. 
This would include realistic residual wavefront errors after an XAO system that will limit the achievable raw contrast to $\sim10^{-5}$ \citep{guyon2018extreme}, while simultaneously creating an incoherent speckle ground that reduces the effectivity of CDI.  
Also, more realistic telescope apertures and coronagraphs, such as the standard Lyot coronagraph and Vector Vortex Coronagraph \citep{mawet2009optical}, need to be {included}.
{Furthermore, recent studies have shown that there are also limitations coming from DM(s) location(s) as well as DM actuator number and quantization errors \citep{beaulieu2017high, beaulieu2020high, ruane2020microelectromechanical}. 
Therefore, we need to add more realistic DM models in future simulations. 
Finally, CDI was studied without any noise sources present other than wavefront aberrations.}
Additional simulations that include these noise sources are needed to accurately predict the gain in contrast that CDI could realistically offer. \\ 

\indent Thus far, we have not discussed how the PESCC Lyot mask with RH polarizer would actually be implemented. 
Unfortunately, it is not as trivial as putting a wire grid polarizer on a substrate in front or behind the RH.
The main problem is the coherence length $L$ of light, which is defined by \citep{wolf2007introduction} {as}: 
\begin{equation}
L = \frac{\lambda^2}{\Delta \lambda}.
\end{equation}
If the optical path difference between the main beam and the beam propagating through the RH polarizer exceeds $L$, then the RH beam becomes incoherent and will not interfere.  
This means that the PESCC would then lose its wavefront sensing capabilities. 
Supposing $\lambda=1550$ nm and $\Delta \lambda = 50$ nm, then $L$ would only be 48 $\mu$m, which sets very tight requirements on the thickness of the RH polarizer. 
For an initial lab demonstration with a laser source, a film or wire grid RH polarizer would be sufficient, as the coherence length of lasers sources is much longer than several meters. 
The implementation in an actual instrument would need other solutions as the bandwidth would otherwise be unacceptably small. 
We envision two possible solutions: 1) reflective Lyot stops with the RH polarizer also operating in reflection; and 2) a simultaneously lithographically etched wire grid polarizer and Lyot stop on thin a glass substrate that covers the entire pupil. 
The former solution will probably be affected by a lower quality of the reflected beam from the polarizer \citep{baur2003new} and is not easily implemented in existing systems as they generally don't have reflective Lyot stops.
Therefore, the latter solution is more appealing as it could be more easily implemented. 
The pixelated polarizer technology offered by Moxtek\footnote{\url{https://moxtek.com/optics-product/pixelated-polarizer/}} looks especially promising, but it will have to be investigated more closely to determine the feasibility of its application. \\

\indent {The PESCC is a variant of the fast-modulated SCC (FMSCC; \citealt{martinez2019fast}), which temporally modulates the RH. 
Here, we put into perspective how the PESCC stands in relation to the FMSCC. 
The two wavefront sensors bear great similarities and part of the simulations presented in this work apply to both (photon noise sensitivity, differential aberrations, and spectral resolution). 
The main difference is the domain in which the reference beam is encoded: polarization for the PESCC and temporal for the FMSCC. 
Both domains have their advantages and disadvantages. 
The PESCC is a completely static solution, important for situations when observing faint targets through the turbulent atmosphere and when moving parts are avoided to prevent vibrations. 
As an added bonus, polarimetry is almost automatically enabled; this is discussed in more detail below.  
However, it comes at the cost of added optics with their own set of requirements (e.g., coherence length, leakage, differential aberrations) and sensitivity to instrumental polarization. 
The FMSCC is optically more easily implemented, as it only adds mechanics to dynamically block the RH in synchronization with camera exposure and readout. 
This makes it suitable for situations in which the number of optics needs to be minimized, for example, in space-based observatories.  
However, in situations when the camera exposure time to get sufficient signal-to-noise becomes longer than the timescale for which the evolving aberrations can be considered frozen, the FMSCC is less applicable. 
The temporal RH modulation might also induce vibrations that can affect the overall system performance. }\\ 

\indent Although the PESCC offers {a factor of 3.5 in spectral bandwidth improvement} compared to the SCC, it still does not encompass an entire broadband photometric band.
Therefore, further improvements are desirable as a broader bandwidth will improve the S/N . 
To increase the bandwidth of the SCC the multi-reference SCC (MRSCC: \citealt{delorme2016focal}) was introduced. 
The MRSCC has additional RHs, placed at different clocking angles, and has shown to reach high contrasts in broad wavelength ranges.   
Similarly, we can introduce the multi-reference PESCC (MRPESCC), which would be very similar to the MRSCC, but with polarizers in each RH. 
These additional RHs would generate fringes in different directions, which enables more accurate broadband wavefront sensing.   
The polarizers in the RHs could be orientated differently, making them sensitive to electric field estimates of opposite polarization states and, therefore, possibly enabling the MRPESCC to measure polarization aberrations \citep{breckinridge2015polarization}. 
Another solution for increasing the bandwidth is via numerical monochromatization of the broadband image \citep{huijts2020broadband}.
In this method, the wavelength scaling of the PSF is inverted by a vector-matrix multiplication, with the vector the flattened broadband image, and the matrix the inverse of the monochromatic image to broadband image mapping.  
The monochromatized image could then be used for wavefront sensing. \\ 

\indent {Starlight is unpolarized to a very high degree, but when it is reflected by an exoplanet, it becomes polarized.}
{Polarization differential imaging (PDI) separates polarized light from unpolarized light, making it a useful tool for discriminating between a planet and mere speckles.}
As the PESCC requires a polarizing beamsplitter, a natural, additional, post-processing method that could be {used} is the PDI, especially for longer integration times, where CDI would have trouble removing the incoherent AO speckle halo; in such a case, the PDI could help remove {this unpolarized structure.}
Using a fast polarization modulator to freeze the atmosphere, subsequent images in one of the channels could directly be subtracted, similar to a {single-beam} polarimeter. 
If we want to combine the CDI and PDI, it has to be investigated whether, following a CDI step on the two polarization channels, it would be possible to directly subtract them (similar to a {dual-beam} polarimeter) or would subsequent images in one channel be subtracted after polarization modulation. 
The latter example would also require a fast polarization modulator ($\sim$100 - 1000 Hz) to ``freeze" the atmosphere. 
An additional advantage of a polarization modulator right after the modified Lyot stop is that the effect of differential aberrations can be minimized.  
A polarization modulator can exchange the two beams between the channels, {that is} the polarization state with reference beam will be in channel 2 instead of channel 1. 
The differential aberration will now flip its sign because two beams now incur the other aberrations. 
Combining two measurements of the {$\Delta \text{OTF,}$}  which had a beam exchange in between them, in time will cancel the detrimental effect of the differential aberrations. 
Similar techniques are used with dual-beam polarimeters to reach high polarimetric sensitivity \citep{snik2013astronomical}. 
In any case, this could be a unique integration of a coronagraph with WFSC, CDI, and PDI. \\ 

\indent The integration of the PESCC in current ground-based high-contrast imaging systems such as Subaru/SCExAO and VLT/SPHERE could be relatively simple. 
Both systems already have focal-plane coronagraphs (\citealt{jovanovic2015subaru}; \citealt{beuzit2019sphere}) and polarizing beamsplitters (\citealt{lozi2019new}; \citealt{de2020polarimetric}) in place. 
Therefore, the only upgrade required would be the modified Lyot stop with RH and polarizer, which in a minimally invasive way could offer substantial gains in terms of focal-plane wavefront sensing and control. 
In the introduction, we do not discuss space-based systems, however, the simulations show that the PESCC could be applicable to them. 
This is even more relevant with regard to space-based systems, such as the high-contrast imaging system in HabEx \citep{mennesson2016habitable} and LUVOIR \citep{pueyo2017luvoir}, where the optics size is limited as the entire telescope is constrained in weight and volume. 
The PESCC might also serve as a powerful solution for such systems, as we demonstrate in this work. \\

\begin{acknowledgements} 
The author thanks S.Y. Haffert and F. Snik for interesting and helpful discussions during this project, and their comments on the manuscript. 
The author also thanks T. Groff for sharing the optical design of the CHARIS polarimetry mode, and M. Oullet for help with Zemax. 
The research of S.P. Bos leading to these results has received funding from the European Research Council under ERC Starting Grant agreement 678194 (FALCONER).
This research made use of HCIPy, an open-source object-oriented framework written in Python for performing end-to-end simulations of high-contrast imaging instruments \citep{por2018hcipy}.
This research used the following Python libraries: Scipy \citep{jones2014scipy}, Numpy \citep{walt2011numpy}, and Matplotlib \citep{Hunter:2007}.
\end{acknowledgements}

\bibliography{report} 
\bibliographystyle{aa} 

\end{document}